\documentclass[fleqn,usenatbib]{mnras}

\usepackage{newtxtext,newtxmath}

\usepackage[T1]{fontenc}

\DeclareRobustCommand{\VAN}[3]{#2}
\let\VANthebibliography\thebibliography
\def\thebibliography{\DeclareRobustCommand{\VAN}[3]{##3}\VANthebibliography}

\usepackage{graphicx}	
\usepackage{amsmath}	
\usepackage{listings}
\usepackage{xcolor}
\newcommand{\msun}{$M_{\odot}$}
\newcommand{\lcdm}{$\rm{\Lambda}$-CDM}

\newcommand{\pkgname}{\texttt{sdssv\_apogee\_sf}}
\newcommand{\pkgurl}{https://github.com/mlucey/sdssv_apogee_sf}
\newcommand{\pkgdocs}{https://sdssv-apogee-sf.readthedocs.io}

\title[SDSS-V APOGEE Selection Function and Mock]{Mapping the Milky Way with SDSS-V: Galactic Genesis Selection Function and Mock Catalog}

\author[Lucey et al.]{
Madeline Lucey,$^{1}$
Zoe Hackshaw,$^{2}$
Robyn Sanderson,$^{1}$
Ilija Medan,$^{3}$  
Alexander~P.~Ji,$^{4,5,6}$ 
Jonathan C.  Bird,$^{7}$
\newauthor
Joel R. Brownstein,$^{8}$
Julie Imig,$^{9}$
Dhanesh Krishnarao,$^{10}$
Juna A. Kollmeier,$^{11}$
Richard R. Lane,$^{12}$
\newauthor
Jianhui Lian,$^{13}$
Andrew K. Saydjari,$^{14}$
and Andrew Tkachenko$^{15}$
\\
$^{1}$Department of Physics \& Astronomy, University of Pennsylvania, 209 S 33rd Street, Philadelphia, PA 19104, USA\\
$^{2}$Department of Astronomy, The University of Texas at Austin, 2515 Speedway Boulevard Austin, TX 78712, USA \\
$^{3}$Canadian Institute for Theoretical Astrophysics, University of Toronto, Toronto, ON M5S-98H, Canada\\
$^{4}$Department of Astronomy \& Astrophysics, University of Chicago, 5640 S. Ellis Avenue, Chicago, IL 60637, USA\\
$^{5}$Kavli Institute for Cosmological Physics, University of Chicago, 5640 S Ellis Avenue, Chicago, IL 60637, USA \\
$^{6}$NSF-Simons AI Institute for the Sky (SkAI), 172 E. Chestnut St., Chicago, IL 60611, USA \\
$^{7}$Department of Physics and Astronomy, Vanderbilt University, 6301 Stevenson Circle, Nashville, TN 37235, USA \\
$^{8}$Department of Physics and Astronomy, University of Utah, 270 S. 1400 E. \#E2108, Salt Lake City, UT 84112, USA\\
$^{9}$Space Telescope Science Institute, 3700 San Martin Drive, Baltimore, MD 21218, USA\\
$^{10}$Department of Physics, Colorado College, 14 East Cache la Poudre St., Colorado Springs, CO, 80903, USA
ORCID iD	0000-0002-7955-7359\\
$^{11}$The Observatories of the Carnegie Institution for Science, 813 Santa Barbara Street, Pasadena, CA 91101, USA\\
$^{12}$Universidad Bernardo O’Higgins, Centro de investigación en Astronomía, Facultad de Ingeniería, Ciencia y Tecnología, Av. Viel 1497, Santiago, 8370993, Chile\\
$^{13}$South-Western Institute for Astronomy Research, Yunnan University, Kunming, Yunnan 650091, People’s Republic of China\\
$^{14}$Department of Astronomy, New Mexico State University, Las Cruces, NM 88003, USA\\
$^{15}$Institute of Astronomy, KU Leuven, Celestijnenlaan 200D, B-3001 Leuven, Belgium
}

\date{Accepted XXX. Received YYY; in original form ZZZ}

\pubyear{\the\year{}}

\begin{document}
\label{firstpage}
\pagerange{\pageref{firstpage}--\pageref{lastpage}}
\maketitle

\begin{abstract}

Robust chemodynamic mapping of the Milky Way requires accounting for the selection function of the survey used, the probability that a given star is included in the final catalog as a function of its observable properties. We derive the empirical selection function of the SDSS-V Galactic Genesis survey, the largest Milky Way Mapper program targeting red giant stars for Galactic structure studies. Treating the Galactic Genesis sample as a subsample of its 2MASS parent catalog, we model the selection function as a function of sky position, apparent H magnitude, and G-H color using a Bayesian binomial estimator, computing both a mean and maximum a posteriori (MAP) version. The resulting selection function reflects the survey's targeting strategy and observing history, including the transition from the pencil-beam SDSS-IV footprint to the fuller sky coverage of SDSS-V and the prioritization of the Magellanic Clouds. Using this selection function with the FIRE-2 Milky Way-mass simulations, we construct mock Galactic Genesis catalogs built on existing mock Gaia DR3 catalogs, injecting realistic photometric noise and applying the selection function via a per-star Bernoulli accept-reject procedure. We demonstrate the mock catalogs' utility with an example application to azimuthal metallicity variations in the disk. Furthermore, we release a Python package allowing users to recompute the selection function for their own samples and science cases. The package, selection function and mock catalogs will be updated and made available as the survey continues, extending the tools needed to reach the full scientific potential of SDSS-V Galactic Genesis.

\end{abstract}




\section{Introduction}

The Milky Way (MW) offers a uniquely detailed case study for understanding galaxy formation and evolution. With billions of individually resolved stars whose chemical and kinematic properties encode the Galaxy's formation history, it remains an unparalleled laboratory for galactic archaeology \citep{Grebel1999,Feast2000,Freeman2002,Sandage2006,Catchpole2016}.

The rise of large-scale surveys have enabled a revolution in chemodynamic studies of the MW. The \textit{Gaia} survey has provided parallaxes and proper motions for billions of stars \citep{Gaia2016, gaiasummary2018, Gaia2023}. Large spectroscopic surveys, e.g., RAVE \citep{Steinmetz2006,Steinmetz2020a}, LAMOST \citep{Deng2012,Zhao2012}, SDSS-III/IV/V \citep{Eisenstein2011,Blanton2017,Kollmeier2026}, and GALAH \citep{De_silva2015,Martell2017},  provide complementary radial velocities and chemical abundances.

These surveys have given rise to numerous studies mapping the correlations between positions, velocities and chemistry of stars in the MW. One of the most fundamental and informative correlations is the distribution of [$\alpha$/Fe]-[Fe/H] as a function of galactocentric cylindrical radius ($R$) and height ($Z$), which constrains the role of various processes governing the galactic disk, including chemical evolution and radial migration \citep{Anders2014, Hayden2015,Gaia_cartography}. More recently azimuthal variations in mean chemical abundances have been characterized and theorized to trace dynamical process, like bar-driven spiral density waves or moving groups \citep{Eilers2022, Poggio2022, Hawkins2023,Filion2023,Hackshaw2024}. Moving beyond the Galactic disk, numerous works have focused on the halo in order to trace the MW's accretion history \citep[e.g.,][]{Carollo2007,Nissen2010,Lee2017,Han2022,Chandra2025,Lucey2026}.

The inferences drawn from these studies depend on how well the underlying samples represent the true stellar population, so it is essential to account for their incompleteness when mapping the MW. Each survey has their own targeting strategy, observational constraints, data processing pipeline, etc. which can all impact exactly which stars are included in the final released catalogs. Furthermore, depending on the specific science case, samples are further down-selected from the catalogs, typically including a quality cut, or identifying a certain stellar type. If not modeled carefully, these processes can introduce biases into the analysis. By defining the selection function, which gives the probability a given star is included in the sample as a function of relevant observational properties, these biases can be properly accounted for \citep{Bovy2016d,Mints2019, Everall2020, Rix2021}.

Since selection functions are crucial for Galactic chemical cartography, they are already modeled for at least some part of the data for many of the large spectroscopic surveys, including SDSS-III/IV/V \citep{Bovy2012b,Bovy2014,Nidever2014,Cantat-Gaudin2024,Medan2026}, RAVE \citep{Wojno2017}, and LAMOST \citep{Chen2018}. For APOGEE, the main survey targets were drawn from 2MASS \citep{Zasowski2013,Zasowski2017}, so the empirical selection function can be estimated as the ratio of observed stars to available 2MASS sources in each field and color-magnitude bin \citep{Cantat-Gaudin2024}. Similarly, SDSS-V BOSS data can mostly be treated as a subset of \textit{Gaia} \citep{Medan2026}. These empirical selection function models are  projections on the sky, and can be converted into effective selection functions as a function of distance using a 3D dust map and a stellar population model drawn from isochrones \citep{Bovy2016d,Cantat-Gaudin2024,Imig2025,Medan2026}. As the \textit{Gaia} survey has provided pivotal data for Galactic structure modeling, there has been huge effort to quantify its selection function \citep{Bovy2017,Boubert2020a,Boubert2020b,Boubert2021,Everall2021,Rybizki2020,Rybizki2021,Everall2022,Cantat-Gaudin2023,Castro-Ginard2023}. For example, \citet{Rybizki2020} provided all-sky magnitude limit maps for \textit{Gaia} DR2 subsets, \citet{Rybizki2021} derived the selection function of the radial velocity sample from its completeness relative to the full \textit{Gaia} catalog, and \citet{Cantat-Gaudin2023} built an empirical model of the \textit{Gaia} DR3 source catalog selection function in terms of $G$ magnitude and sky position. Selection functions have also been determined for specific stellar types like RR Lyrae, where completeness is estimated empirically by comparing independent catalogs as a function of sky position and distance \citep{Mateu2020}. This was updated for \textit{Gaia} DR3 and a combined \textit{Gaia}+PS1+ASAS-SN catalog by \citet{Mateu2024}.

One method commonly used to vet, and utilize selection functions is to create mock catalogs. There are generally two approaches to this. One samples empirically derived density distributions and typically assumes dynamical equilibrium \citep{Robin2003,Girardi2005,Juric2008,Rybizki2018,Rybizki2020}. However, in order to include dynamical disequilibria beyond analytical models, the other approach samples star particles from MW-mass cosmological simulations \citep{Sharma2011,Lowing2015}. This method had been used to make mock \textit{Gaia} catalogs \citep{Grand2018,Sanderson2020,Nguyen2024}, as well as a mock SDSS-IV catalog \citep{Nikakhtar2021}. 

As one of the main goals of the SDSS-V survey is to map the MW \citep{Kollmeier2026}, determining the selection function of the survey is crucial. We specifically focus on the Galactic Genesis (GG) survey (Bird et al. in prep), which uses the 2MASS \citep{Skrutskie2006} and \textit{Gaia} DR2 \citep{gaiasummary2018} surveys for targeting as described in Section \ref{sec:sdssv}. The FIRE-2 MW-mass simulation, which are used to make a mock GG catalog, is presented in Section \ref{sec:sims}.  In Section \ref{sec:sf}, we derive the selection function empirically, and in Section \ref{sec:mock} create and verify the mock catalog.  Last, Section \ref{sec:sum} summarizes this work and the conclusions.

\section{SDSS-V Milky Way Mapper} \label{sec:sdssv}
The SDSS-V survey is divided between three mappers: The Black Hole Mapper (BHM), The Local Volume Mapper (LVM) and the Milky Way Mapper (MWM). One of the main goals of the MWM is to study the Galaxy's formation and evolutionary history using chemical cartography and galactic dynamics. This is achieved by collecting millions of optical and infrared spectra using the SDSS telescope \citep{Gunn2006} in the northern hemisphere and the du Pont telescope \citep{Bowen1973} in the southern hemisphere. In this work, we focus on the infrared data from the APOGEE instruments, which have wavelength coverage of $1.51-1.70~\mu m$ and $R\approx22,500$ \citep{Wilson2019}. These spectra are the primary data for Galactic chemodynamic studies in the MWM survey.

The targeting of APOGEE data in SDSS-V is complex due to a wide variety of science goals in the survey. These goals are divided between programs, each with its own target selection procedure. For more details of all the programs and their targeting strategy we refer the reader to \citet{Kollmeier2026} and the SDSS website\footnote{Targeting information is available at the following address: \url{https://www.sdss.org/dr20/targeting/overview/} Please make note to update the url for the data release relevant for your purpose.}. Furthermore, a number of programs have released papers describing their target selection in more details, including the Magellanic Genesis program \citep{Nidever2026} and the Young Stellar Objects program \citep{Kounkel2023}. However, the largest program in MWM is the GG survey, which we focus on in this work as it is specifically designed for population and MW structure studies. 

Once targets are selected by a science program, there are still a number of steps that reduce the number of stars to what actually makes it into the final data release catalog. Firstly, there is a complex procedure for which targets actually get assigned to a fiber in the observing program, taking into account priority and fiber positioning constraints \citep{Medan2025,Blanton2026}. As all of the observations for BHM and MWM are planned at once, there are many competing programs which impact the fiber assignment in complicated ways. Weather and technical issues may cause deviations from this plan that can also impact which stars get observed and included in the final catalog. Finally, there may be steps in the data analysis pipeline that result in certain stars not being included in specific catalogs for a variety of reasons, including data artifacts, or poor calibrations. These can all be difficult to track which motivates an empirical estimate of the selection function, which is informed by the targeting strategy, rather than trying to model directly each of these effects. 

\subsection{The Galactic Genesis Survey}

The main goal of the GG survey is to obtain millions of APOGEE spectra of red giant stars across the sky with minimal selection biases, in order to study the MW's history through the chemodynamics of its stars (Bird et al. in prep). Therefore, the targeting strategy is intentionally simple, using 2MASS and \textit{Gaia} DR2 photometry. However, SDSS-V includes data from SDSS-IV which used different targeting strategy and fiber plug plates \citep{Zasowski2017,Beaton2021,Santana2021}. Furthermore, small changes in targeting selection criteria occurred after SDSS-V DR19. The targeting flags\footnote{\url{https://www.sdss.org/dr20/targeting/flags}} in the final data products delivered by the \texttt{astra} pipeline (Casey et al. in prep) label any data that pass the targeting selection criteria for a given program agnostic to the targeting strategy used when it was observed.  Therefore, we include SDSS-IV data that meets the SDSS-V DR19 targeting criteria : 
\begin{equation} \label{eqn:targ}
     H < 11 \ \text{and} \ \big[ (G-H) > 5 \
    \text{or no } G \text{ reported in \textit{Gaia} DR2} \big] 
\end{equation} 
\citep{sdssdr18,sdssdr19}.
For data releases beyond DR20, this criteria will expand to:
\begin{align} \label{eqn:targv2}
    H < 11 \ \text{and} \ \big[ &(G-H) > 5 \ \text{or no } G \text{ reported in \textit{Gaia} DR2}  \  \\
    &\text{or } 
    \varpi < 100*10^{0.2(-2.5-H)}
    \big] \nonumber
\end{align}
Therefore, in data releases beyond DR20, the SDSS-V GG survey may include more data from SDSS-IV because of this additional or statement. In this work, we use the latest public APOGEE release which is DR19, since there was no new APOGEE data released in DR20 \citep{sdssdr20}.

APOGEE also has a nominal bright limit of $H>7$ but, using fiber offsets, a handful of stars with $4.5<H\leq7$ have been observed. This selection targets very red stars, including stars too red to be observed by \textit{Gaia}, as GG is primarily interested in cool giants embedded in the dusty Galactic disk, particularly in the highly extincted inner Galaxy. The parallax cut is intended to select giant stars and corresponds to $H_{abs}<-2.5$. It is important to note that the ``or'' statements mean that not every star in the sample meets all three criteria within the bracket, but at least meets one of them. For example, stars with $G-H<5$ may be included if they satisfy the parallax-$H$ criterion. 

In this work, we focus primarily on results from the ASPCAP pipeline \citep{GarciaPerez2015,Holtzman2018,Jonsson2020}, which provides stellar parameters and abundances for APOGEE spectra and is the most widely used pipeline for Galactic studies \citep[e.g.,][]{Hayden2015,Feuillet2019,Queiroz2020,imig2023,Griffith2024}. We use this final catalog (\texttt{astraAllStarASPCAP-0.6.0.fits}) rather than an earlier-stage product, such as the list of all observed stars, because we aim to capture the full impact of the data analysis pipeline on the selection function. These are the products actually used in scientific analyses.

\section{FIRE-2 Cosmological Zoom-in Simulations} \label{sec:sims}

We create the mock SDSS-V catalog by employing the publicly available\footnote{\url{https://flathub.flatironinstitute.org/fire}} \citep{Wetzel2023} \textit{Latte} suite of seven isolated MW-mass galaxies \citep{Wetzel2016}, together with the \textit{ELVIS} suite of three Local Group-like pairs of comparable mass \citep{Garrison-Kimmel2019}. Each simulation adopts the FIRE-2 physics model \citep{Hopkins2018b} and is evolved with the GIZMO\footnote{\url{http://www.tapir.caltech.edu/~phopkins/Site/GIZMO.html}} gravity and hydrodynamics code, run in meshless finite-mass (MFM) mode \citep{Hopkins2015}. Full implementation details are given in these references; here we summarize the properties most relevant to this work.
All runs assume a flat \lcdm\ cosmology consistent with \citet{Planck2014}. Except for m12w, the \textit{Latte} suite adopts $\Omega_m$ = 0.272, $\Omega_b$ = 0.0455, $\sigma_8$ = 0.807, $n_s$ = 0.961, $h$ = 0.702. The \textit{ELVIS} pairs Thelma \& Louise and Romulus \& Remus retain the cosmology of the original dark-matter-only \textit{ELVIS} suite: $\Omega_m$ = 0.266, $\Omega_b$ = 0.0449, $\sigma_8$ = 0.801, $n_s$ = 0.963, $h$ = 0.71.  The remaining pair, Romeo \& Juliet, along with m12w, instead follow the updated parameters \citet{Planck2020}: $\Omega_m$ = 0.31, $\Omega_b$ = 0.048, $\sigma_8$ = 0.82, $n_s$ = 0.97, $h$ = 0.68.

The mass distribution of simulated galaxies is known to depend sensitively on how stellar feedback is treated \citep{Pontzen2014,Lazar2020}. In FIRE-2, star formation is self-regulated through a combination of stellar winds, radiation pressure from young stars, Type II and Type Ia supernovae, photoelectric heating, and photoionization. Star formation occurs above a gas density threshold of $n_{SF}>1000~\rm{cm^{-3}}$, with feedback rates, luminosities, energies, and mass-loss rates taken directly from stellar evolution models \citep[STARBURST99;][]{Leitherer1999}. Across the 13 simulated galaxies, $z=0$ dark matter halo masses span $\rm{M_{200m} = 1-2.1 \times 10^{12}~M_{\odot}}$ \citep{Samuel2020,Barry2023}. Baryonic resolution differs by suite: initial stellar particle masses are 7070 \msun\ for \textit{Latte} and 3500 \msun\ for the higher-resolution \textit{ELVIS} runs, with star particle softening lengths of $\approx$4 pc and dark matter softening of $\approx$40 pc. Corresponding dark matter particle masses are 35,000 \msun\ for \textit{Latte} and 19,000 \msun\ for \textit{ELVIS}. These choices give numerical convergence at the $\approx$100 pc scale \citep{Wetzel2016}, well below the $\lesssim$950 pc disk scale heights these simulations resolve \citep{Yu2021}. Numerical effects should therefore not strongly impact our analysis of the stellar halo, which focuses on scales $>1$ kpc.

FIRE-2 galaxies provide useful mock catalogs of the MW because they broadly match several key MW properties, including the stellar halo mass fraction \citep{Sanderson2018}, the presence of a metal-rich in-situ halo component \citep{Bonaca2017}, and the radial and vertical structure of the stellar disk \citep{Ma2017,Sanderson2020,Bellardini2021,McCluskey2024}. Simulated satellite populations around both the MW and M31 analogs are likewise consistent with observations \citep{Wetzel2016,Samuel2020,Garrison-Kimmel2019b,Panithanpaisal2021,Cunningham2021}. Some tensions with MW data remain, however: although FIRE-2 reproduces the observed stellar masses and number of tidal streams, \citet{Shipp2023} found that stream orbital properties are not well reproduced. \textit{Latte} disks also tend to assemble later than observational estimates of the MW disk age suggest \citep{McCluskey2024}. While these differences should be kept in mind, they do not invalidate the utility of the FIRE-2 simulations for constructing mock catalogs.

\section{Selection Function} \label{sec:sf}

\begin{figure*}
    \centering
    \includegraphics[width=\linewidth]{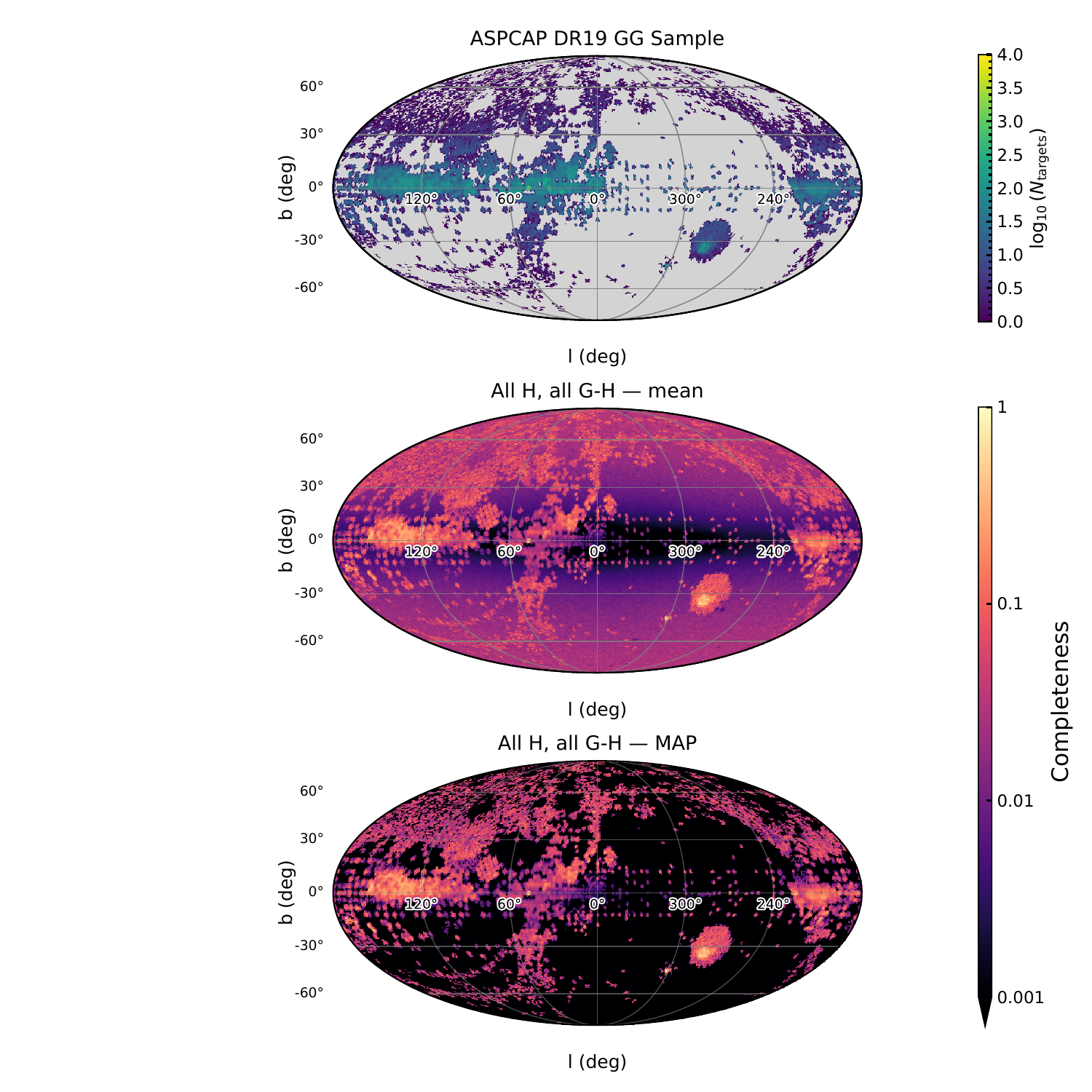}
    \caption{ The sky distribution of the GG DR19 data (top panel) and the corresponding selection function in galactic coordinates with the Galactic center at (0,0). The middle panel shows the mean of the posterior version of the selection function marginalized over all the H and G-H bins, while the bottom panel shows the MAP version. For both the middle and bottom panels, we can see the selection function is highest in regions where the GG sample has high density, as expected. The strongest difference between the mean and MAP versions is also visible here, where the MAP version gives a selection function value of 0 in parts of the sky where there are no GG observations, while the mean version gives small but nonzero values that depend on the total number of stars observed in 2MASS at that sky positions.  }
    \label{fig:skysf}
\end{figure*}

\begin{figure*}
    \centering
    \includegraphics[width=\linewidth]{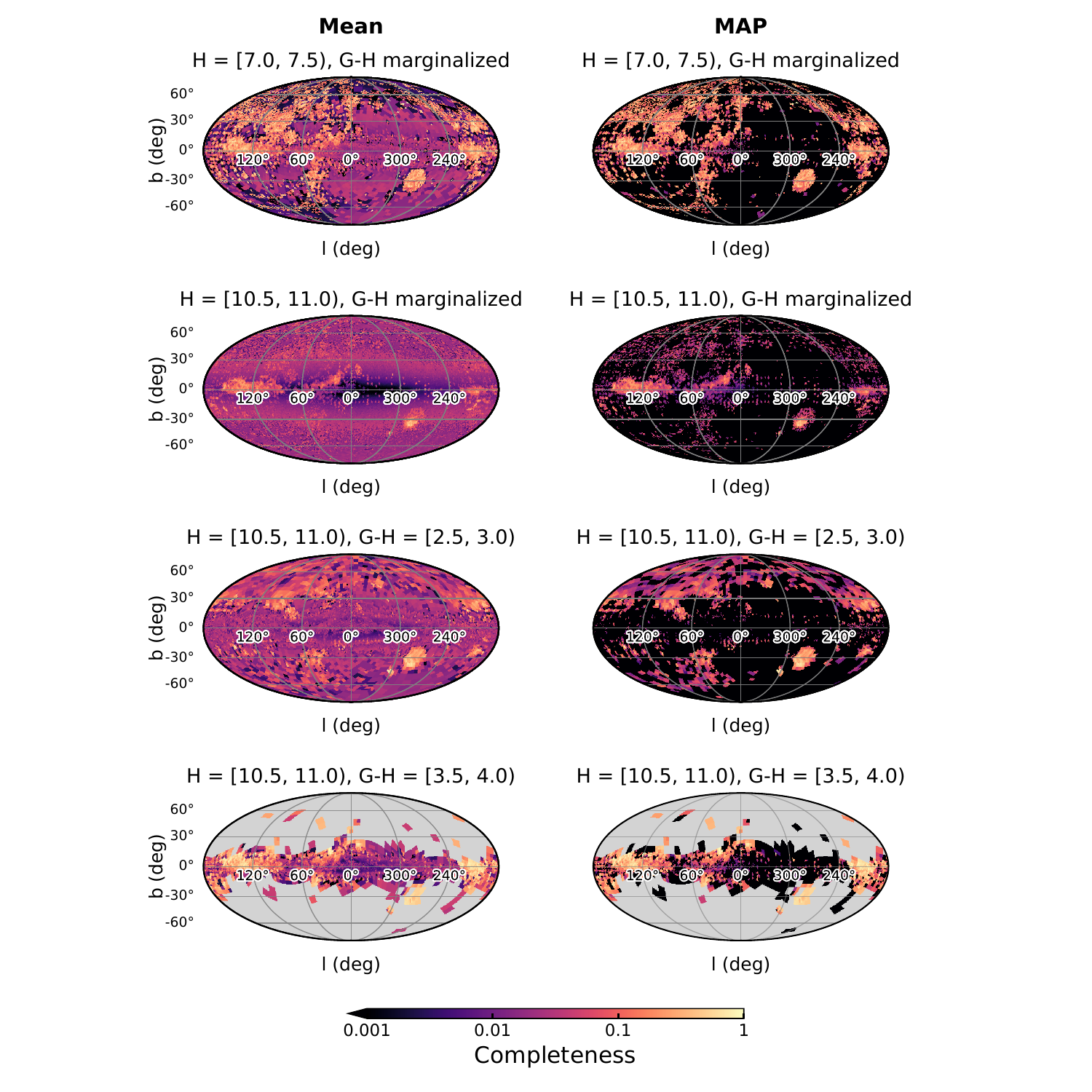}
    \caption{ The  sky distribution of the mean (left column) and MAP (right column) versions of the selection function in galactic coordinates. The top two rows show the selection function evaluated at a specific H bins, marginalized over the color axis. The top row shows the bright end of APOGEE observations with $7\leq H<7.5$ while the second row shows the faint end with $10.5\leq H <11$. The bottom rows show again the faint end but now evaluated at specific color bins. The third row from the top ($2.5\leq G-H<3$) shows one of the most populated bins  that is still blue enough to contain stars that are in the halo and not substantial reddened. The bottom row shows the selection function for the most populated bin ($3.5\leq G-H<4$). Pixels that do not have over 20 2MASS targets are shown in grey and are not used to calculate the selection function. Most pixels in the bottom row are grey because stars are typically only that red with the high levels of dust extinction in the Galactic plane.   }
    \label{fig:skysfcolor}
\end{figure*}

\begin{figure*}
    \centering
    \includegraphics[width=\linewidth]{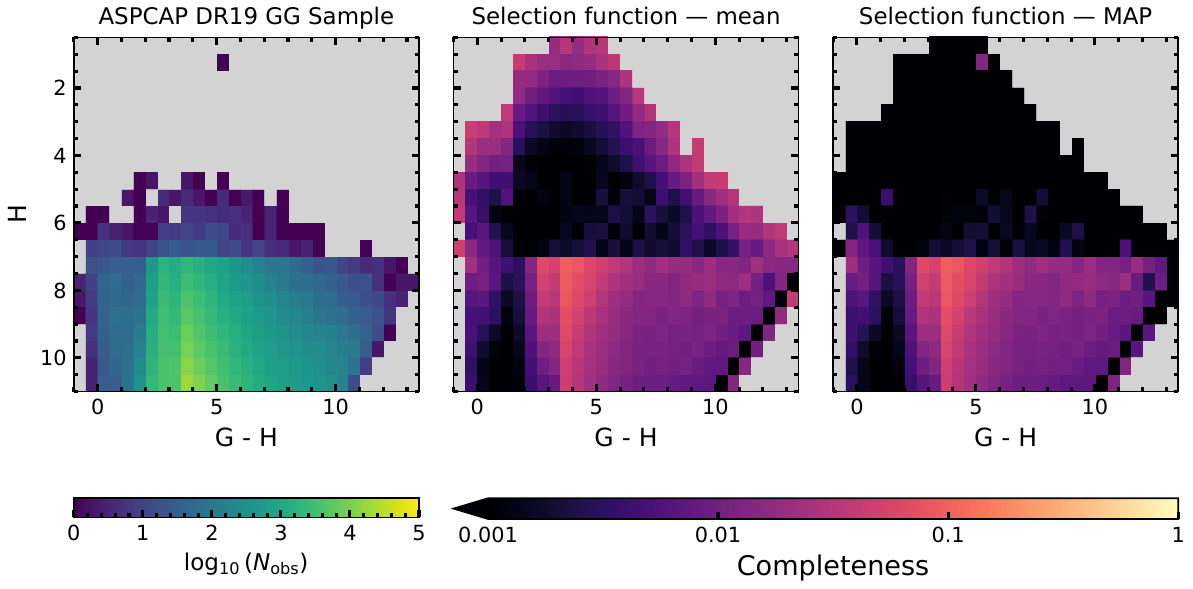}
    \caption{ The observed apparent color magnitude diagram (left panel) compared to the selection function. The middle panel shows the mean version of the selection function, while the right panel shows the MAP version. The bounds are set by the observed distribution, and we do not calculate the selection function in H, G-H bins with fewer than 20 2MASS stars to avoid biases from low number statistics. In general the selection function is not a strong function of $H, G-H$, besides the nominal magnitude range of the survey ($7<H<11$), but the sky position is a much bigger effect.    }
    \label{fig:cmdsf}
\end{figure*}

As described in Section \ref{sec:sdssv}, the selection of targets in SDSS-V is a complex process. The data collection, reduction and analysis process further impacts the final sample available for scientific inferences. As these effects are difficult to track, it is best to instead empirically model the selection function. To achieve this we follow a procedure similar to \citet{Cantat-Gaudin2024} and \citet{Medan2026}, where we estimate the completeness as a function of sky position, and apparent magnitude ($H$) with an optional additional color axis ($G-H$). Therefore, our selection function assumes that stars are exchangeable within each cell of sky position, $H$, and $G-H$, so each cell value is the probability, averaged over the cell, that a 2MASS source reaches the final ASPCAP catalog, even though individual probabilities vary with factors such as fiber collisions and competing targets. Our spatial cells range from $\sim$1$^\circ$ across at $N_{\rm side}=64$ to $\sim$7$^\circ$ across at $N_{\rm side}=8$ where counts are sparse, comparable to the $3^\circ$ diameter of an SDSS-V field, so each cell averages over the small-scale effects of fiber collisions. This within-cell variation does not bias population-level inference provided it is independent of the stellar properties being inferred, which holds for fiber assignment because it depends on sky position, carton priority, and robot reach rather than on stellar chemistry or kinematics. However, this is  only valid within the possible parameter range of the ASPCAP catalog given by the model grid \citep{Jonsson2020}. For example, ASPCAP fails for stars with [M/H]$\,<-2.5$, which are therefore absent from the catalog regardless of their position, magnitude, or color. 

As SDSS-V primarily uses the 2MASS survey \citep{Cutri2003,Skrutskie2006} for APOGEE targeting, we can treat it as a subsample of the 2MASS catalog and use the following equation:
\begin{equation}
\mathcal{S}_\mathcal{C}^{\rm subsample}(\mathbf{q}) = \mathcal{S}_\mathcal{C}(\mathbf{q} | \mathbf{q} {\rm\ in\  parent}) \cdot \mathcal{S}_\mathcal{C}^{\rm parent}(\mathbf{q})
\end{equation}
where $\mathcal{S}_\mathcal{C}(\mathbf{q} | \mathbf{q} {\rm\ in\  parent})$ is the probability a source with attributes $\mathbf{q}=\{l,b,H,...\}$ will be in the subsample given that it is in the parent catalog. $\mathcal{S}_\mathcal{C}^{\rm parent}(\mathbf{q})$ is the corresponding probability that a given source will be in the parent catalog. 

The 2MASS point source catalog is 99\% complete at H=15 in high-latitude regions where source confusion is negligible\footnote{\url{https://irsa.ipac.caltech.edu/data/2MASS/docs/releases/allsky/doc/sec2_2b.html}}. Towards the Galactic center, the completeness limit occurs at significantly brighter magnitudes. While the completeness here is not measured as robustly, the analysis available indicates that it is safe to assume the 2MASS point source catalog is complete up to H=12\footnote{\url{https://irsa.ipac.caltech.edu/data/2MASS/docs/supplementary/sec6_7jmc/analysis.html}}. If this assumption over estimates the completeness of 2MASS the SDSS-V selection function will also be over -estimated. However, there are limited sources in SDSS-V with H>12 and the GG survey specifically only targets stars with H<11. Therefore, the impact on the overall calculation of the selection function should be limited and we assume  $\mathcal{S}_\mathcal{C}^{\rm parent}(\mathbf{q}) =1$ within our $\mathbf{q}$ bounds.

For sources with attributes $\mathbf{q}$, we define $p(\mathbf{q})$ as the probability that a single source is selected into the sub-sample. Given $n$ sources in the parent and $k$ sources in the subsample catalog sharing attributes $\mathbf{q}$, the likelihood of $p(\mathbf{q})$
\begin{equation} \label{eqn:like}
    \mathcal{L}(p) \propto  p^k (1-p)^{n-k}.
\end{equation}

Using an uninformative uniform prior, the posterior can be described as a $\mathrm{Beta}(k+1, n-k+1)$. We use two methods of deriving a point estimate for $\mathcal{S}_\mathcal{C}^{\rm parent}(\mathbf{q}) = p(\mathbf{q})$ from this posterior. 
One method is the mean or expected value of the posterior which is given by:
\begin{equation} \label{eqn:exp}
\mathrm{E}(p) = \frac{k+1}{n+2}  
\end{equation}
We also use the maximum a posteriori (MAP) which is given by:
\begin{equation} \label{eqn:emap}
p_{MAP}(\mathbf{q}) = \frac{k}{n}  
\end{equation}

The expectation value is typically used as the estimate of $\mathcal{S}_\mathcal{C}^{\rm parent}(\mathbf{q})$ \citep[e.g.,][]{Castro-Ginard2023,Medan2026}. However, the expectation value is never zero even in bins with zero observed targets. In order to create a mock catalog which accurately recreates the sky and magnitude distribution of the observed DR19 GG sample, we require bins with zero targets to have a selection function value of zero, so  mock stars in regions of the sky and magnitude space that SDSS did not observe do not leak into the catalog. The MAP value achieves this (see Section \ref{sec:mock} and Figures \ref{fig:mocksky} and \ref{fig:mockkiel}). On the other hand, however, the MAP value does not distinguish a cell with a genuinely zero selection probability from a sparsely sampled cell in which no stars happened to be observed.

We calculate the expectation values and MAP in bins of $\mathbf{q}$. Given that the targeting strategy for the GG survey contains the color term $G-H$ (see Equation \ref{eqn:targ}), we also include it as an optional additional dimension of $\mathbf{q}$. However, there are sources in 2MASS that do not have a reported \textit{Gaia} DR2 $G$ magnitude, likely mostly because of dust extinction. For those sources, we instead calculate their minimum G-H, given their H magnitude and the \textit{Gaia} DR2 faint magnitude limit of G=21 \citep{gaiasummary2018}. We note that only 2.5\%  of 2MASS sources ($\approx$ 350,000 stars) with $4.5<H<11$ do not have a match in \textit{Gaia}. Therefore, this $G-H$ approximation will not have a large effect on our selection function calculation, but it is still important to include in the calculation of $n$ so that the completeness is not overestimated. 

The GG targeting strategy also includes stars that do not have a reported \textit{Gaia} DR2 G magnitude. When the $G-H$ dimension is included in  $\mathbf{q}$, we exclude these stars from the numerator ($k$), so in this case we are specifically calculating the selection function of the GG sample with \textit{Gaia} photometry. These stars make up a small fraction of the GG sample, only 5,967 out of 235,219 stars (2.5\%) in DR19. Their missing \textit{Gaia} photometry in SDSS-V catalogs may arise in part from differences between cross-match chains or from bookkeeping issues, rather than a true non-detection. Therefore, assigning these stars G = 21 could spuriously inflate the selection function at red G-H, so we exclude them from the color selection function.

When the $G-H$ term is not included in $\mathbf{q}$, we simply do not use \textit{Gaia} data at all and the selection function is computed only as a function of sky positions and $H$ magnitude, rather than marginalizing over $G-H$. Therefore, stars missing \textit{Gaia} $G$ magnitudes can easily be included in both numerator ($k$) and denominator ($n$).

We nominally use bins of 0.5 mag in both H and G-H. We note that in the accompanying Python package \pkgname\ \citep{sfpackage} (see Appendix \ref{app:pack}), users can specify their own binning. The binning in sky coordinates ($\ell, b$) is performed using HEALPix \citep{Gorski2005,Zonca2019}. We use adaptive binning to decrease the resolution until a minimum of 20 2MASS objects per bin (n$\geq$20). With this limit, the statistical uncertainty on the selection probability in any cell is at most $\sim$0.1, and a cell with no observed stars still constrains the selection probability to $\lesssim$0.14 at 95\% confidence. In sparsely populated $(H, G-H)$ bins, the adaptive scheme must coarsen the HEALPix resolution until each cell contains at least 20 2MASS stars, which can wash out real small-scale structure in the survey footprint. To recover this structure, we assume that the small-scale angular pattern of the completeness varies only weakly with $H$ over broad magnitude ranges, so that it can be measured in wider $H$ bins and shared among the native bins they contain. We therefore merge the native $H$ bins into two such bins ($H < 7$ and $7 \le H < 11$), whose larger star counts allow the completeness to be estimated at finer angular resolution. Within each native cell, we then redistribute the original completeness across the finer sub-pixels in proportion to the completeness of the corresponding merged bin, normalized so that the expected number of observed stars in the cell is exactly preserved. The hybrid therefore leaves the selection function unchanged on the original coarse scale and only adds angular detail within each cell. In high stellar density regions, we achieve fine HEALPix resolution with nside=64, which corresponds to angular resolution of $0.9^{\circ}$ and a physical resolution of $\approx$ 130 pc at the Galactic center. In the lowest density regions, we set a minimum bin size of nside=8 with angular resolution of $7.33^{\circ}$ and a physical resolution of $\approx$ 1 kpc at the Galactic center. 

In Figure \ref{fig:skysf}, we show the observed DR19 GG survey sky distribution (top panel) compared to the empirical selection function. The middle panel shows the mean of the posterior version of the selection function, while the bottom panel shows the MAP version. In general, the selection function has the highest completeness in regions of the sky where the GG sample has its highest density. However, this is not always the case, especially in the inner Galaxy, where the GG sample has many targets, but the total underlying number of stars there is very high so that the completeness is still quite low. Certain areas of the sky have more complete coverage because of the observing strategy, including the Magellanic clouds which were prioritized in SDSS-IV. We also see the remnants of the SDSS-IV pencil-beam footprint, where SDSS-V is now filling in with full-sky coverage. The mean and MAP versions of the selection function are generally very similar except for one important difference. In the regions of sky where there are no observed GG targets, the mean selection function value is small but nonzero (because of the +1 term in the numerator), while the MAP version report exactly zero. This has important consequences when applying the selection function for creating a mock catalog as shown in Section \ref{sec:mock}.

We further investigate the color and magnitude dependence of the selection function in Figure \ref{fig:skysfcolor}. The left column shows the mean version of the selection function, while the right column shows the MAP version. The top panels shows the bright end of APOGEE observations with $7\leq H <7.5$, while the second row shows the faint end with $10 \leq H < 11$. The bottom rows show the selection function evaluated at the faint end with two different $G-H$ bins. Specifically, we show one of the most densely populated bins that still has full sky coverage in the third row from the top ($2.5 \leq G-H < 3$ ).  The $H, G-H$ bin with the highest density of targets is $10.5 \leq H < 11$ and $3.5 \leq G-H < 4$ and is shown in the bottom row. The selection function is only calculated here along the disk and the Magellanic clouds, because of the lack of 2MASS targets elsewhere in the galaxy that are sufficiently red. These stars only exist in areas with significant dust extinction.

Figure \ref{fig:cmdsf} shows the observed distribution of $H, G-H$ (leftmost panel) compared to the empirical selection function. The middle and right panels show the selection function as a function of $H, G-H$ marginalized across the sky with the mean version in the middle and the MAP version in the right panel. The bright and faint limits ($7<H<11$) of the APOGEE GG survey are clearly visible here. There are a few stars with $H<7$, observed with fiber offsets, but they are a small fraction of the total sample. The grey bins do not have a selection function value because they do not meet the criteria of having at least 20 stars in the 2MASS catalog. Again the difference between the mean and MAP versions of the catalog is visible here where we see bins with no GG observations have selection function value of exactly zero in the MAP version, but have small nonzero values in the mean version.

\section{Mock Catalogs} \label{sec:mock}

\begin{figure*}
    \centering
    \includegraphics[width=\linewidth]{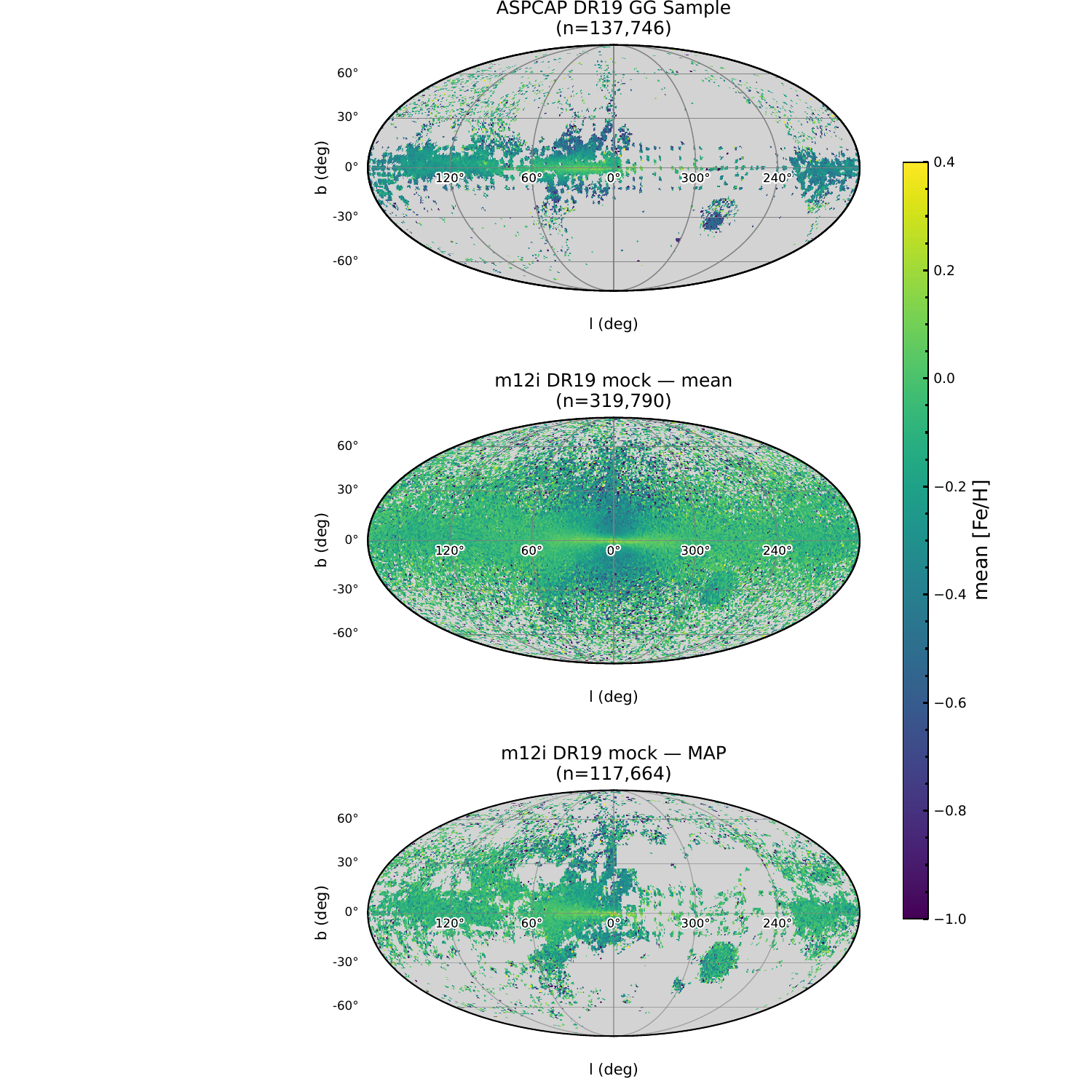}
    \caption{ The  sky distribution of the GG DR19 data (top panel) compared to the mock catalogs in galactic coordinates with the Galactic center at (0,0) and colored by mean metallicity. The middle panel shows the mock calculated with the mean version of the selection function, while the bottom panel shows the same for the MAP version. Since the MAP selection function is exactly zero in bins with no stars, the mock better reflects the true distribution of the data, with additional spread due to the spatial resolution of the selection function calculation.  }
    \label{fig:mocksky}
\end{figure*}

\begin{figure*}
    \centering
    \includegraphics[width=\linewidth]{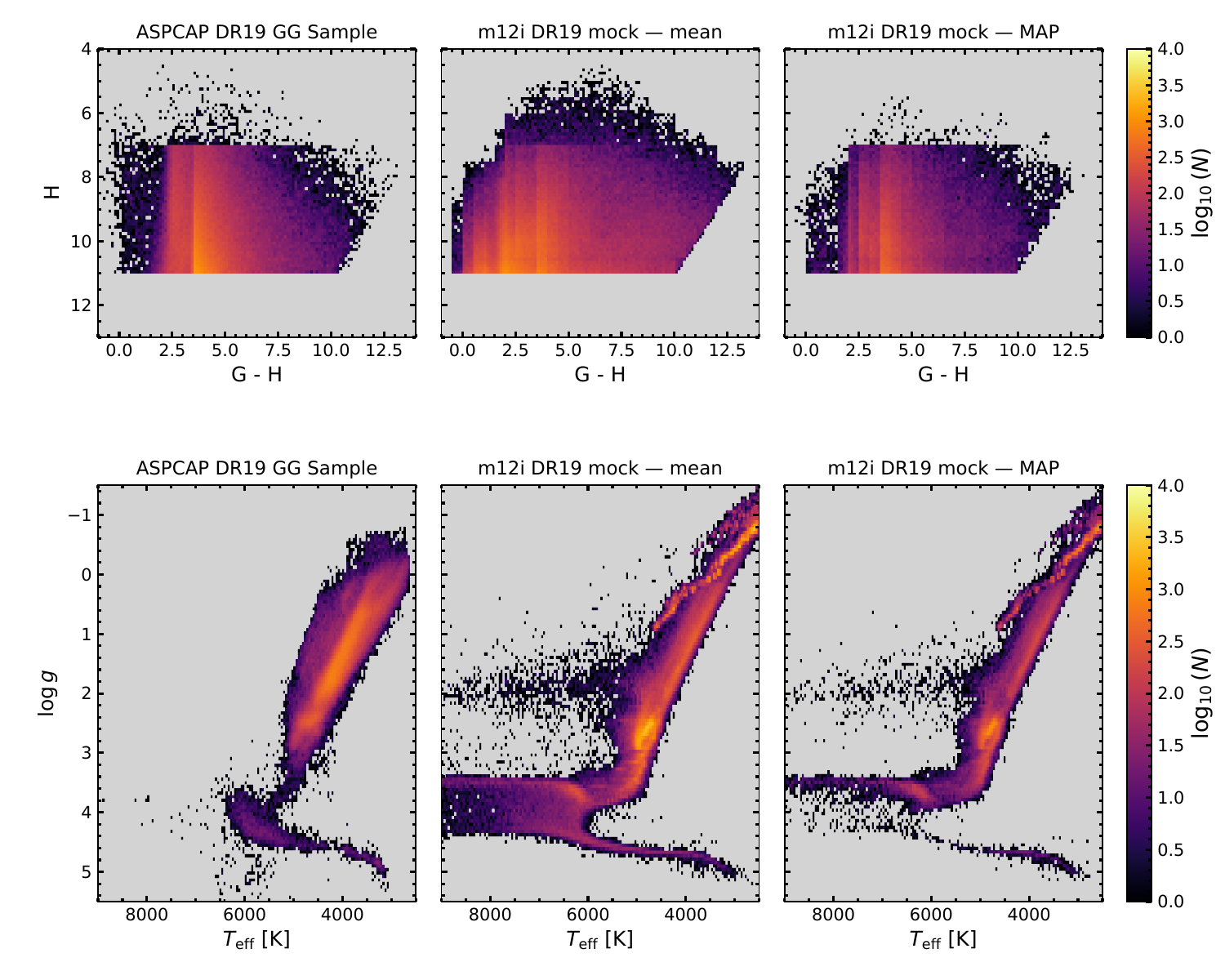}
    \caption{ The observed apparent color magnitude diagram (top left panel) compared to the mock catalogs with (middle top panel) and without the prior (right top panel). The bottom panels show the corresponding Kiel diagrams. The prior-free mock more closely resembles the observed H and G-H distribution but differences still exist because of the resolution of the binning resolution of the completeness calculation. Differences in the Kiel diagrams between the observed and mock catalogs are driven by imperfect isochrone models as well as limitations in the ASPCAP analysis pipeline. }
    \label{fig:mockkiel}
\end{figure*}

Using this empirical selection function, we build mock catalogs designed to reproduce the SDSS-V APOGEE survey. Mock catalogs of this kind are valuable tools for comparing simulations or theoretical models directly to MW data \citep{Robin2003,Sharma2011,Sanderson2020,Grand2018}. By imposing the same selection effects present in the real survey, they let us test which physical quantities are actually measurable given the data in hand, and identify observational selection effects that could bias the resulting scientific conclusions \citep{Nandakumar2017,Mints2019,Everall2020}. 

As we are interested in the SDSS-V APOGEE sample that is in \textit{Gaia} DR3, we use the mock \textit{Gaia} DR3 catalogs from \citet{Nguyen2024} as the starting point for our SDSS-V mocks. We refer the reader to \citet{Nguyen2024}, for the full details of the process to create these catalogs. In short, these catalogs are built from the Latte suite of FIRE-2 cosmological zoom-in baryonic simulations of MW-mass galaxies \citep{Wetzel2016}, part of the Feedback In Realistic Environments (FIRE) simulation project, which were run using the Gizmo gravity plus hydrodynamics code in meshless finite-mass (MFM) mode \citep{Hopkins2015} and the FIRE-2 physics model \citep{Hopkins2018}. They use the \textsc{Ananke} framework to transform star particles from the simulations into synthetic stars with \textit{Gaia} DR3-like parallaxes and photometry \citep{Sanderson2020,Thob2024}. Specifically, \textsc{Ananke} uses the \textsc{Galaxia} code \citep{Sharma2011} with the \citet{Kroupa2001} initial mass function (IMF). \citet{Nguyen2024} specifically uses the updated \textit{Gaia} DR3 Padova CMD v3.6\footnote{\url{http://stev.oapd.inaf.it/cgi-bin/cmd}} isochrones. Since these simulations do not directly model the formation of dust grains, the extinction is assumed to be proportional to the metal-weighted column density of hydrogen along the line of sight \citep{Rachford2009,Sanderson2020}. The detailed \textit{Gaia} DR3 empirical selection function \citep{Cantat-Gaudin2023} is not applied and instead a simple G cut (3<G<21) based on the \textit{Gaia} DR3 limiting magnitudes is employed. Therefore, this mock overselects stars, as it does not account for the lower completeness in regions affected by crowding and the \textit{Gaia} scanning law. However, this does not impact its usefulness as a starting point for our mock catalog where we will apply the SDSS-V empirical selection function.

The first step to build the SDSS-V APOGEE catalogs is to infer 2MASS magnitudes for the mock \textit{Gaia} DR3 catalogs. We use the \textsc{Galaxia} code to calculate the 2MASS magnitudes using Padova isochrones \citep{Girardi2002,Cohen2003,Marigo2008}. We then calculate the dust extinction using $A_J/A_V, A_H/A_V, A_{Ks}/A_V$ from \citet{Wang2019}. We calculate two versions of the true apparent magnitudes. The first uses the E(B-V) values reported in \citet{Nguyen2024}, which are based on the distribution of metal-rich gas in the simulations, so that the star and dust distributions are self-consistent. This self-consistency is most important for regions with ongoing star formation, where the dense gas and young stars are most strongly correlated \citep{Sanderson2020}. The second version instead calculates the E(B-V) from MW dust maps, specifically using the \textsc{Bayestar} package \citep{Green2018} using the DECaPS2-based map for the southern hemisphere \citep{Zucker2025} and the map from \citet{Green2019} for the northern hemisphere with Pan-STARRS coverage \citep{Chambers2016}. In the MW dust case we also calculate the extinction in the \textit{Gaia} G band using the \textsc{dustapprox} package \citep{Fouesneau2025} which calculates the $k_G=A_G/A_0$ as a function of $A_0$ (the monochromatic extinction at 550 nm) and effective temperature ($T_{eff}$), consistent with \citet{Nguyen2024}. Since the dust in both the MW dust map and the self-consistent extinction estimation primarily lies on the Galactic plane, either version will give similar results for older stellar populations.

To inject realistic photometric noise into synthetic 2MASS/APOGEE catalog, we characterize the magnitude-dependent uncertainty reported by the 2MASS Point Source Catalog (PSC) empirically, rather than assuming an analytic error curve. For each of the three 2MASS bands ($J$, $H$, $K_s$), we query the full PSC and bin every source by its catalog magnitude in that band using 0.1 mag bins spanning $-5 \le m \le 17$, wide enough to cover the saturated-bright to noise-floor-faint ends of all three bands. Within each bin we compute the source count and the mean $\mu(m)$ and standard deviation $s(m)$ of that band's combined photometric-uncertainty, reducing the empirical uncertainty distribution in each bin to a Gaussian approximation, $\sigma \sim \mathcal{N}(\mu(m), s(m))$. Bins containing fewer than 20 sources are flagged as unreliable and discarded, so small-number statistics do not drive the model at the sparsely populated bright and faint extremes. See Appendix \ref{sec:noise} for a plot of the noise models. For the recalculated \textit{Gaia} G magnitude we also recalculate the $Gaia$ DR3 G mag uncertainty using the \textsc{pygaia} uncertainty model \citep{Riello2021}. To simulate photometry for the mock catalogs, the uncertainty ($\sigma$) then perturbs the true magnitude via $m_{\rm obs} = m_{\rm true} + \mathcal{N}(0,\sigma)$.

Each mock star's position and (noisy) $H$, $G-H$ are passed to the selection function query, returning the probability that a real 2MASS star at that position and magnitude would have been targeted and observed by APOGEE. This probability is then applied per star as a Bernoulli accept/reject draw (retained with probability $S$, discarded otherwise), producing a synthetic "APOGEE-observed" mock catalog that carries the same selection effects, including its stochastic incompleteness pattern, as the real data. The resulting mock catalog can then be compared directly to the observed APOGEE sample without needing to reproduce the underlying target-selection algorithm itself.

In addition to the stellar parameters ($T_{eff}$ and log $g$) the FIRE-2 simulations have a number of elemental abundances that are in common with APOGEE. These elements are Fe, O, Mg, Si, Ca, S, C and N. We include these  elemental abundances for our mock stars with injected APOGEE/ASPCAP-like noise. Similar to the procedure for 2MASS magnitudes, we approximate the distribution of APOGEEE signal-to-noise (SNR) as a function of H band magnitude with a Gaussian mean and standard deviation and use this to probabilistically assign an SNR to each mock star. We again follow a similar procedure for deriving an empirical relation between APOGEE SNR to ASPCAP stellar parameters ($T_{eff}$ and log $g$) as well as the chemical abundances, [Fe/H], [O/H], [Mg/H], [Si/H], [Ca/H], [S/H], [C/H] and [N/H]. While this procedure reliable recreates ASPCAP-like uncertainties, it is well-known that DR19 ASPCAP uncertainties are underestimated and do not take into account systematics \citep{Meszaros2025}.

In Figure \ref{fig:mocksky} we show the sky distribution colored by mean metallicity of the DR19 ASPCAP GG sample (top panel) compared to the mock catalogs calculated with the mean  (middle panel) and the MAP (bottom panel) versions of the selection function. Although some fields are more complete, much of the DR19 sky distribution is dominated by the pencil beam design of SDSS-IV. SDSS-V will continue to fill this in more in future releases. The mock catalog from the MAP version of the selection function better matches the sky distribution of the real data. This is because the MAP selection function is exactly zero in pixels with zero observed stars, but for the mean version the minimum value is 3.95$\times10^{-5}$ because the numerator is never zero (see Equation \ref{eqn:exp}). The MAP version of the mock is still more dispersed across the sky than the observed sample because of the spatial resolution in the selection function estimate. Higher spatial resolution could be achieved at the cost of coarsening the magnitude binning or decreasing the minimum number of 2MASS observations per bin and therefore increasing the noise in the completeness estimate. 

Overall, the mock contains more stars out of the plane than the DR19 data. Specifically, only 5.7\% of the DR19 stars have $|b|>15^{\circ}$, while this represents 23.5\% of the MAP mock. This is partially driven by the lower spatial resolution that can be achieved in the selection function at these higher latitudes where 2MASS stellar counts are more sparse. Furthermore, there is evidence that the FIRE-2 cosmological simulations have more massive stellar halos than the MW \citep{Lucey2026c}. This may also contribute to why the average [Fe/H] at high galactic latitudes ($|b|>15^{\circ}$) is +0.35 dex more metal-rich in the MAP mock than the DR19 data. The average [Fe/H] within the galactic plane  ($|b|<5^{\circ}$) for the MAP mock and DR19 data agree to within 0.06 dex.

SDSS-IV completed an extensive mapping of the LMC and SMC with APOGEE \citep{Nidever2020, Hasselquist2021}. A lot of this data also satisfies the target selection of the GG program, leading to the MC coverage as a significant feature in the DR19 ASPCAP GG sky distribution (see the top panel of Figure \ref{fig:mocksky}). While we recreate this feature in the sky distribution, it is important to note that we do not actually ensure the presence of an MC-like satellites at those sky positions in the FIRE-2 simulations  or resulting mocks. Therefore, it is essentially just a random halo field which is significantly more metal-rich than the MCs. 

Figure \ref{fig:mockkiel} compares the apparent magnitude and color distribution of the ASPCAP GG DR19 sample (top left panel) to the mock catalogs calculared from the mean (top middle panel) and MAP (top right panel) versions of the selection function. The bottom panels show the corresponding Kiel diagram. There are clear differences between the observed  and the mock catalogs' color magnitude distributions, although the MAP version is more similar to the observed than the mean version. These differences are partially driven by the coarseness of magnitude binning in the selection function estimate (see Figure \ref{fig:cmdsf}).  The mock catalog contains more targets on the edges of the the observed distribution where the selection function estimate is small but not zero. This effect is heightened in the mean version where bins with zero observed targets have a non-zero, but small selection function estimate. 

Overall the $\rm{T_{eff}}$ distribution of stars is much hotter in the mock than the DR19 observed data. For example, 97.7\% of the DR19 data has $\rm{T_{eff}}$<5000 K, while the same is true for only 82.9\% of the MAP mock. Furthermore, there are 4,512 stars in the MAP mock with $\rm{T_{eff}}$>7000 K, while the DR19 data has only 42. The mocks also contain a higher percentage of turn-off/sub-giant stars, relative to the the DR19 data. 17\% of the mock has 3$\leq$ log $g$ $\leq$4, while the same is true for only 1\% of the DR19 data.  

Some of the differences in the Kiel diagram between the observed ASPCAP DR19 GG sample and the mocks are driven by the difference in the present-day star formation rate between the MW and the simulated galaxies; the simulations have a local SFR several times that of the MW \citep{Sanderson2020} resulting in a larger proportion of young stars in the mock catalogs. Additionally, while the targeting is based on H and G-H, the ASPCAP analysis pipeline further limits the  $\rm{T_{eff}}$ and log $g$ range. ASPCAP cannot measure log $g$>5 or log $g$<-0.5. While ASPCAP can handle stars with 2500 K $\leq \rm{T_{eff}} \leq $ 20,000 K, but only for stars with log $g$<3. The hot star range is clipped to $\rm{T_{eff}} \leq $ 8,000 K for stars with 3 $\leq$ log $g \leq$ 1 and further clipped to $\rm{T_{eff}} \leq $ 6,000 K for 1 $\leq$ log $g \leq$ 0 and $\rm{T_{eff}} \leq $ 4,000 K for 0 $\leq$ log $g \leq -0.5$. Furthermore, ASPCAP cannot measure parameters for stars with metallicity $<-2.5$, which may contribute to the lack of blue horizontal branch stars in the observed sample. While ASPCAP may be capable of deriving parameters within the range described above, it can fail more frequently in certain regimes. For example, 63\% of stars with G-H<2 are flagged as bad by ASPCAP, which roughly corresponds to $\rm{T_{eff}} > 6000$ K. The mocks also have sharper features (e.g., red clump) in the Kiel diagrams because the ASPCAP uncertainties are underestimated, including only the statistical contribution and excluding the larger systematic uncertainties \citep[see Appendix \ref{sec:noise};][]{Meszaros2025}.

The evolved stellar populations are also sensitive to the choice of stellar models used to create the isochrones that sample the stellar populations for the mocks \citep[e.g.][]{2013MNRAS.434..488M, 2016ApJ...822...73R,2019MNRAS.485.5666P,2020MNRAS.498.3283P}, and to the choice of models of the most massive stars \citep{2014MNRAS.445.4287T}. These models have been calibrated against observations but only in a limited range of metallicities compared to the range needed to sample all the simulated stellar populations. Any of these can lead to differences with the observed ASPCAP catalog in the regions where the discrepancies are most apparent in Figure \ref{fig:mockkiel}. This specifically drives the strong diagonal feature near the top of the giant branch which are mostly stars with age $\geq$ 10 Gyr. 

\subsection{Azimuthal Variations Example}
\begin{figure*}
    \centering
    \includegraphics[width=\linewidth]{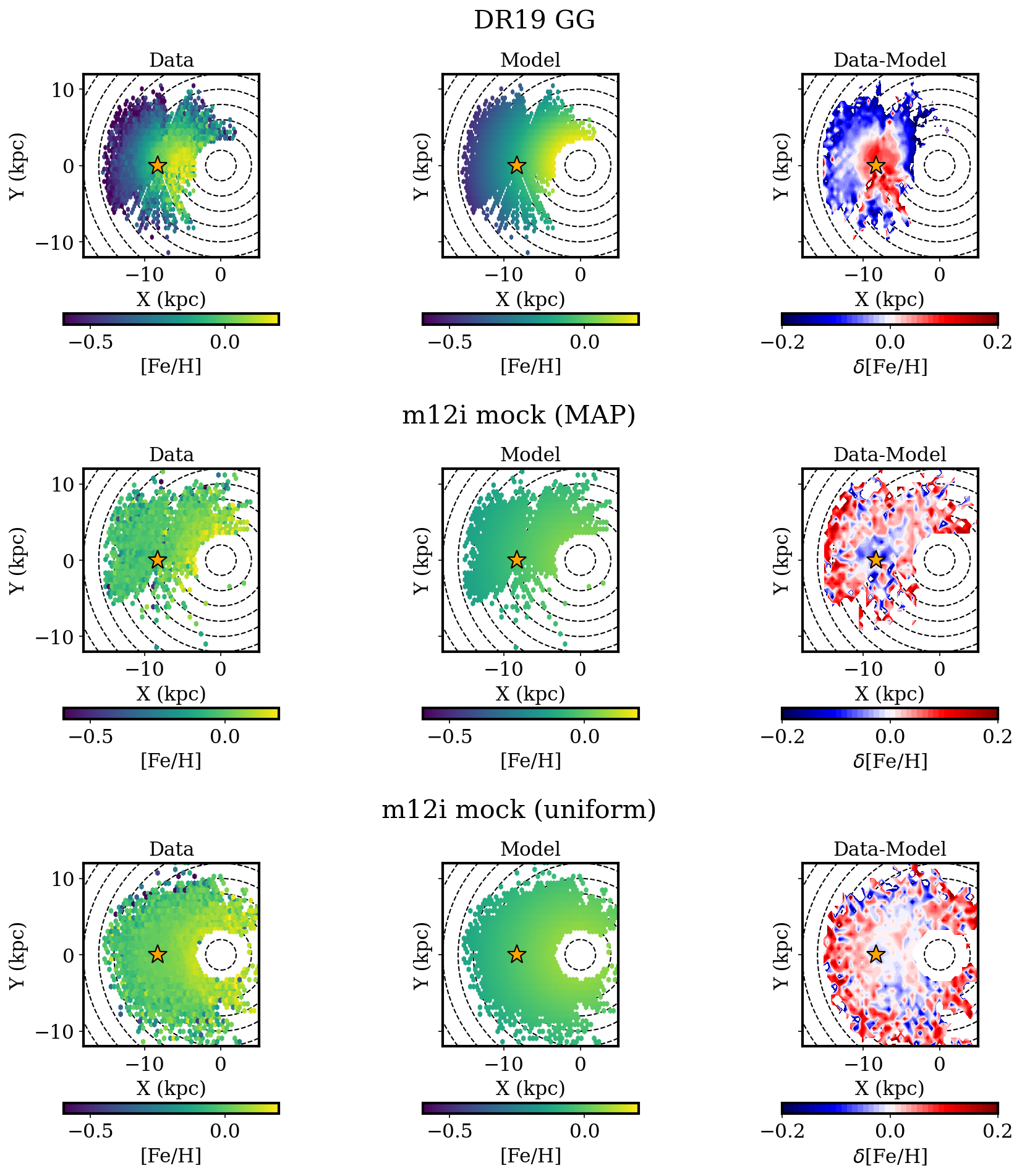}
    \caption{Top row: The azimuthal metallicity variations observed in DR19 with the astroNN GG sample. Middle row: The azimuthal metallicity variations in the m12i MAP mock. Bottom row: The azimuthal metallicity variations in an m12i mock with a uniform selection instead of the GG DR19 selection function. In all rows, the leftmost panel is the planar disk sample colored by metallicity. The center panel is the model of the metallicity gradient, where each star is assigned a metallicity solely based on the negative metallicity gradient in the sample. The rightmost panel subtracts the model metallicity from the observed metallicity, showcasing the azimuthal metallicity variations as $\delta$[Fe/H]. The uniform mock and DR19 MAP mock show substantial differences in the azimuthal substructure, suggesting that the MW's substructure is also changed by selection effects.   }
    \label{fig:azvars}
\end{figure*}

One of the most prominent observed chemical features in the MW disk is the negative radial metallicity gradient \citep[an incomprehensive list includes:][]{Mayor1976,Andrievsky2002,Magrini2009,Luck2011,Bergemann2014,Xiang2015,Yan2019,Hawkins2022,imig2023}. This gradient is theorized to be the result of the `inside-out' formation of the Galaxy, by which the inner Galaxy forms first and fast, igniting rapid star formation and increasing the metal content of stars located towards the Galactic center \citep[e.g.,][]{Matteucci1989,Chiappini1997}.

On top of this gradient, azimuthal substructure in metallicity has been reported on the $\sim0.1$ dex-level. Different tracers have exhibited this azimuthal structure, including hot OBA stars \citep{Hawkins2022} in LAMOST, hot giant stars in Gaia \citep{Poggio2022}, and open clusters \citep{Otto2026}. While the origin of these azimuthal metallicity variations likely have dynamical origins \citep{Jurado2026}, it is apparent that they are a key physical feature of the disk. 

To identify if this azimuthal substructure is present in the presented mock catalogs, we follow the methodology put forth in \citet[][hereafter H24]{Hackshaw2024}. H24 used APOGEE DR17 stars to identify a negative radial metallicity gradient of $\Delta$[Fe/H]/$\Delta$R $\sim-0.0678\pm 0.0004$ dex/kpc. They identified azimuthal variations on top of this gradient, with the magnitude of these azimuthal variations correlating with various dynamical parameters. Similarly, we identified cool ($3500 <$ T$_{\text{eff}}$ $< 5000$ K) giant (log $g$ $< 3.6$ dex) stars in the DR19 astroNN GG catalog. We removed any stars with errors greater than 0.1 dex on [Fe/H] and [Mg/H], as well as any stars errors greater than 150 K and 0.3 dex on T$_{\text{eff}}$ and log $g$, respectively. We then selected stars within 0.3 kpc of the Galactic midplane and $3.5 < $R$_{\text{galactocentric}} < 15.5$ kpc to select stars confined to the plane and minimize the effects of the bar in the inner Galaxy \citep[i.e.,][]{Hammersley1994,Wegg2015,Lucey2023}. 

To further investigate the impact of the selection function on the derived azimuthal substructure, we create a mock version that uses a uniform selection which sets the ground truth. Specifically, this selection randomly down samples the \textit{Gaia} DR3 mock with a 5\% acceptance rate. As the \textit{Gaia} mock merely applies a selection based on G band magnitude that is not a function of the sky position or color, this sample gives an unbiased representation of the true underlying disk population. 

Following H24, we identify negative radial metallicity gradients in the resulting DR19 astroNN GG sample and both the m12i mock samples. Azimuthal substructure in $\delta$[Fe/H] is identified in all samples after subtracting the radial gradient off of the sample metallicities. Figure \ref{fig:azvars} shows the gradients, model metallicities, and resulting azimuthal variations in both samples. It is already well established that FIRE-2 galaxies do not reproduce the strength of the MW's radial metallicity gradient \citep{Graf2025}. Here, we confirm that accounting for selection effects does not resolve this discrepancy. However, the azimuthal substructure in the m12i MAP mock differs substantially from that of the uniform m12i mock. This suggests that the selection function also shapes the observed azimuthal structure of the MW and must be accounted for to recover its true substructure.

\section{Summary and Conclusions} \label{sec:sum}

The MW remains an unparalleled laboratory for detailed studies of galaxy formation, and large-scale surveys like Gaia and SDSS-V are essential for mapping its chemodynamic structure. Robust mapping requires accounting for each survey's selection function, the probability that a given star is included in the sample as a function of observational properties. While selection functions have been derived for many major surveys \citep[e.g.,][]{Wojno2017,Chen2018,Cantat-Gaudin2023,Cantat-Gaudin2024}, this had not yet been done for the SDSS-V GG survey, but is a key ingredient to achieving many of the survey's science goals. 

In this work, we calculate the SDSS-V GG selection function as a subsample to its 2MASS parent catalog, which we assume to be complete within the relevant magnitude range of 4<H<11. Based on the targeting strategy, we calculate the selection function as a function of sky position, H and G-H. We calculate two versions, one using the mean of the posterior consistent with previous works \citep[e.g.,][]{Castro-Ginard2023,Medan2026} and one with the MAP. The main difference is that the MAP version has the selection function equal to zero in bins where zero stars are observed while the mean version instead reports a small but nonzero value. 

With this selection function, we also utilize FIRE-2 MW-mass galaxy simulations to create mock DR19 GG ASPCAP catalogs, building upon the \textit{Gaia} DR3 mocks from \citet{Nguyen2024}. We create two versions of the model photometry one with MW dust maps \citep{Green2019, Zucker2025} and another with the dust distribution consistent with the simulation. We inject noise by deriving a 2MASS and ASPCAP noise model based on apparent magnitude. We also provide further mock catalog variations based on whether the mean or MAP version of the selection function is used, as well as using different FIRE-2 simulations and 

We work a short example use case of the mock catalogs by comparing the azimuthal variations between the observed and mock data following the procedure presented in \citep{Hackshaw2024}. This comparison shows that the selection function substantially alters the observed azimuthal substructure, highlighting the importance of modeling it when interpreting the MW's chemical structure. In general, the selection function and corresponding mock catalog provide crucial tools for reaching the full scientific capability of the SDSS-V GG survey. For future data releases (DR21 and DR22), the mock catalogs will be provided as a Value Added Catalog (VAC) and the selection function will be updated and availabile on the corresponding Github package\footnote{\url{https://github.com/mlucey/sdssv_apogee_sf/tree/main}} (see Appendix \ref{app:pack}).

\section*{Software}
This research made use of \textsc{Astropy} \citep{astropy:2013,astropy:2018,astropy:2022}, \textsc{NumPy} \citep{numpy}, \textsc{Matplotlib} \citep{matplotlib}, \textsc{pandas} \citep{pandas:2010,pandas:2020}, \textsc{astroquery} \citep{astroquery}, \textsc{healpy} \citep{Zonca2019,Gorski2005}, and \textsc{h5py} \citep{h5py}. Generative AI (Anthropic's Claude, via Claude Code) was used to help write and debug analysis code and the \texttt{sdssv\_apogee\_sf} package, and edit portions of the text. All AI-assisted output was reviewed, tested, and revised by the authors, who takes full responsibility for the content of this work.

\section*{Acknowledgements}

This material is based upon work supported by the National Science Foundation under Award No. 2303831.
RES gratefully acknowledges travel support from the Simons Foundation under award 1018462.
APJ acknowledges support from NSF grant AST-2510795, the Alfred P. Sloan Research Fellowship, and the University of Chicago’s Research Computing Center.
J.I. gratefully acknowledges support from STScI DDRF project D0001.82536

Funding for the Sloan Digital Sky Survey V has been provided by the Alfred P. Sloan Foundation, the Heising-Simons Foundation, the National Science Foundation, and the Participating Institutions. SDSS acknowledges support and resources from the Center for High-Performance Computing at the University of Utah. SDSS telescopes are located at Apache Point Observatory, funded by the Astrophysical Research Consortium and operated by New Mexico State University, and at Las Campanas Observatory, operated by the Carnegie Institution for Science. The SDSS web site is \url{www.sdss.org}.

SDSS is managed by the Astrophysical Research Consortium for the Participating Institutions of the SDSS Collaboration, including the Carnegie Institution for Science, Chilean National Time Allocation Committee (CNTAC) ratified researchers, Caltech, the Gotham Participation Group, Harvard University, Heidelberg University, The Flatiron Institute, The Johns Hopkins University, L'Ecole polytechnique f\'{e}d\'{e}rale de Lausanne (EPFL), Leibniz-Institut f\"{u}r Astrophysik Potsdam (AIP), Max-Planck-Institut f\"{u}r Astronomie (MPIA Heidelberg), Max-Planck-Institut f\"{u}r Extraterrestrische Physik (MPE), Nanjing University, National Astronomical Observatories of China (NAOC), New Mexico State University, The Ohio State University, Pennsylvania State University, Smithsonian Astrophysical Observatory, Space Telescope Science Institute (STScI), the Stellar Astrophysics Participation Group, Universidad Nacional Aut\'{o}noma de M\'{e}xico, University of Arizona, University of Colorado Boulder, University of Illinois at Urbana-Champaign, University of Toronto, University of Utah, University of Virginia, Yale University, and Yunnan University.  

This work has made use of data from the European Space Agency (ESA) mission
{\it Gaia} (\url{https://www.cosmos.esa.int/gaia}), processed by the {\it Gaia}
Data Processing and Analysis Consortium (DPAC,
\url{https://www.cosmos.esa.int/web/gaia/dpac/consortium}). Funding for the DPAC
has been provided by national institutions, in particular the institutions
participating in the {\it Gaia} Multilateral Agreement.

\section*{Data Availability}
The many ways to access SDSS-V data are described here: \url{https://www.sdss.org/dr20/data_access/}. The FIRE-2\footnote{\url{https://flathub.flatironinstitute.org/fire}} public data releases \citep{Wetzel2023} are available on the Globus platform.



\bibliographystyle{mnras}
\bibliography{bibliography} 



\appendix 
\section{2MASS and ASPCAP Noise Model} \label{sec:noise}

\begin{figure}
    \centering
    \includegraphics[width=\linewidth]{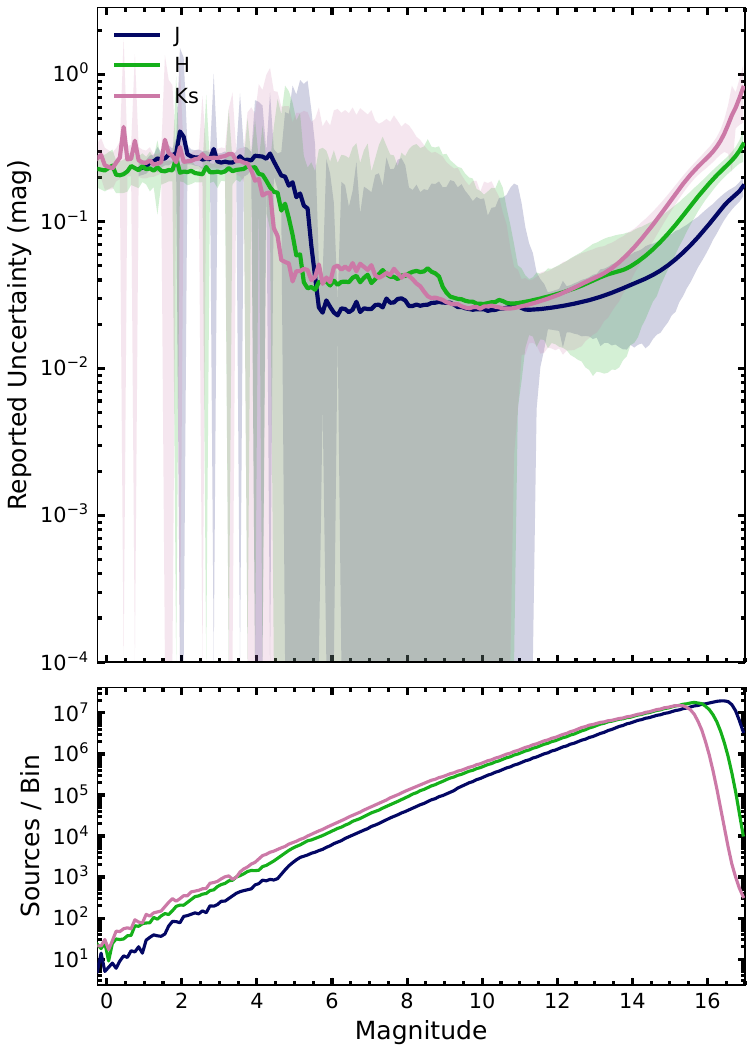}
    \caption{Empirical photometric noise model for the 2MASS Point Source Catalog. The top panel shows the mean reported photometric uncertainty (\texttt{msigcom}) as a function of magnitude in 0.1\,mag bins for the $J$ (blue), $H$ (green), and $K_s$ (pink) bands. Shaded regions show the $\pm1\sigma$ scatter within each bin. Each bin is modeled as a Gaussian with this mean and standard deviation, and mock photometric uncertainties are drawn from it. The bottom panel gives the number of sources per bin. Only bins with at least 20 sources are used, and the magnitude range is cut at 17\,mag, just fainter than the 2MASS detection limit. Beyond this limit, a few spurious or blended sources with inflated uncertainties dominate the sparsely populated bins.} 
    \label{fig:2mass_nosie}
\end{figure}

\begin{figure}
    \centering
    \includegraphics[width=\linewidth]{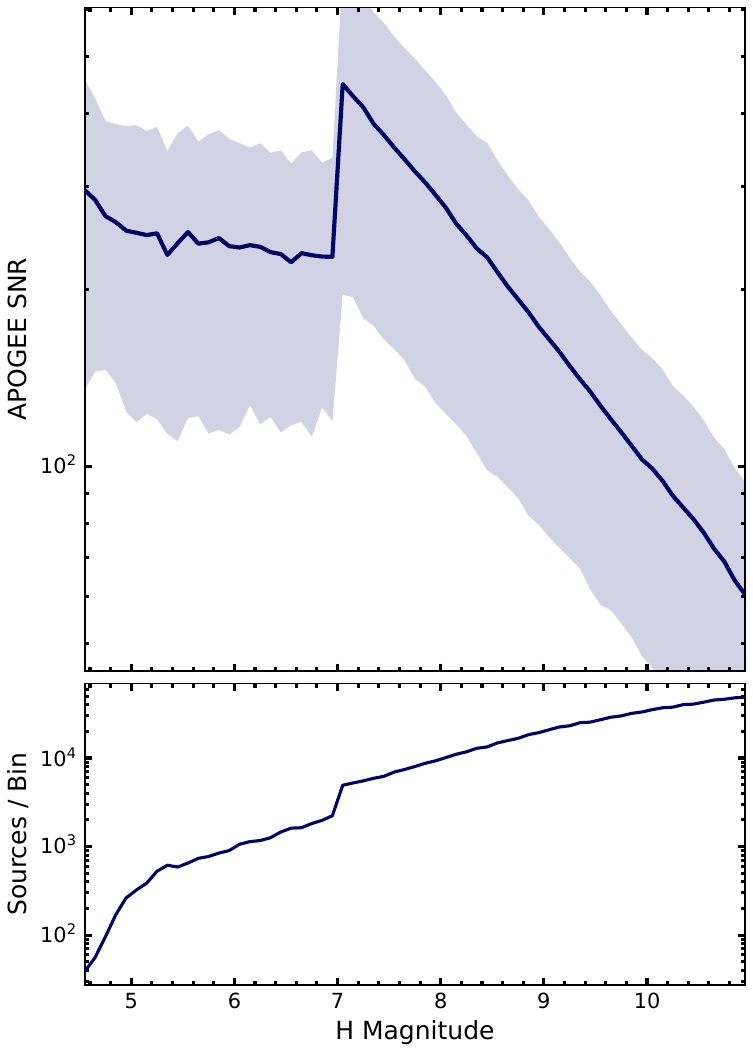}
    \caption{APOGEE spectra signal-to-noise ratio as a function of 2MASS $H$ magnitude. The top panel shows the mean SNR in 0.1\,mag bins (line) with the $\pm1\sigma$ scatter within each bin (shaded). The bottom panel gives the  number of stars per bin. Only bins with at least 20 stars are shown. Each bin is modeled as a Gaussian in SNR, so a mock star's SNR is drawn given its $H$ magnitude. This SNR is then passed to the per-parameter uncertainty relations in Figure~\ref{fig:snr_noise}.}
    \label{fig:apo_snr}
\end{figure}

\begin{figure*}
    \centering
    \includegraphics[width=\linewidth]{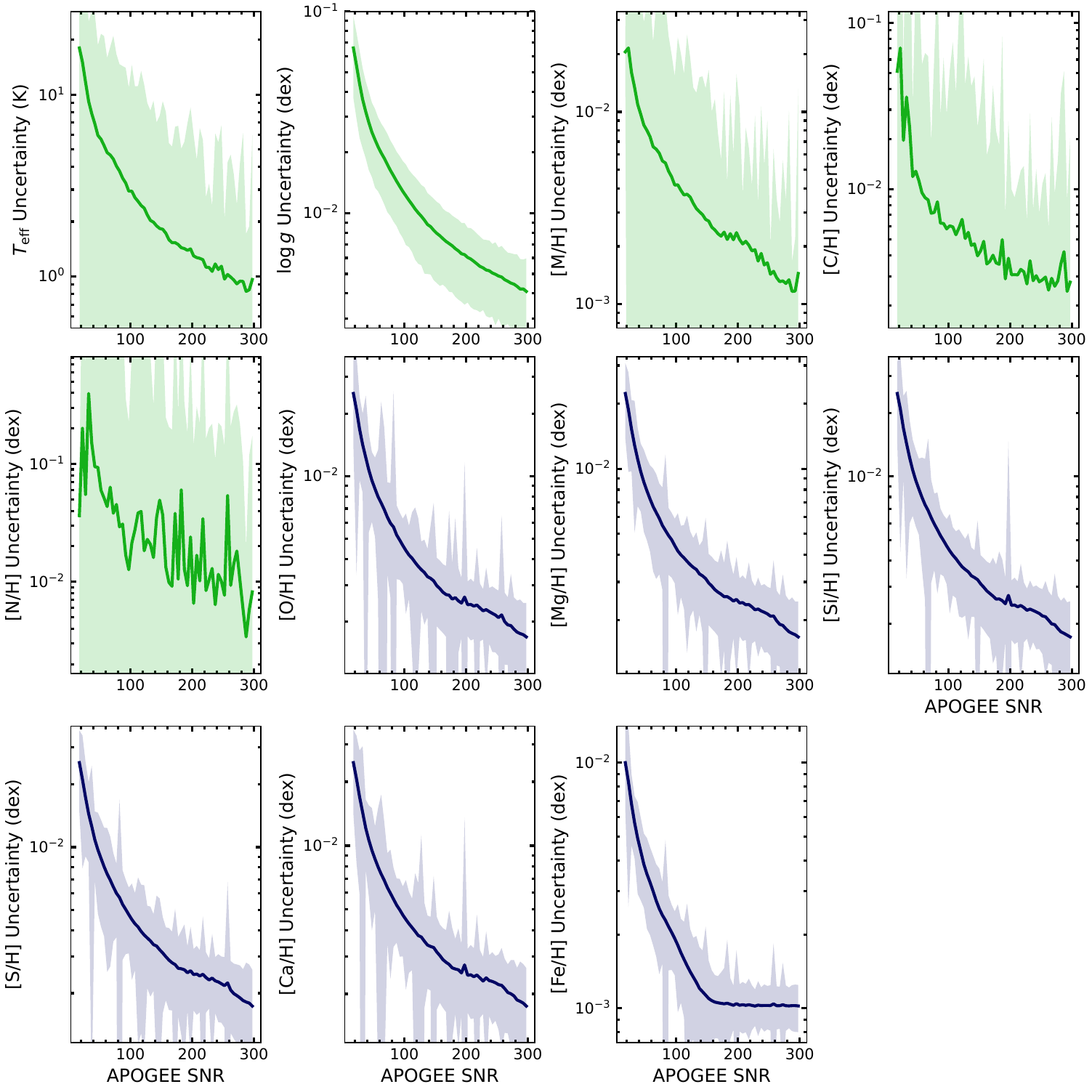}
    \caption{The reported uncertainty on each stellar parameter as a function of APOGEE SNR. Lines show the mean reported uncertainty in bins of $\Delta\mathrm{SNR}=5$ between $\mathrm{SNR}=0$ and 300. Shaded regions show the $\pm1\sigma$ scatter within each bin. $\rm{T_{eff}}$, log $g$, [M/H], [C/H], and [N/H] are shown in green as they are solved for simultaneously in a joint fit using the full spectrum, while the other abundances ([O/H], [Mg/H], [Si/H], [S/H], [Ca/H] and [Fe/H]) are solved for individually in specific wavelength regions. Mock uncertainties are drawn from a Gaussian in each bin, using the SNR drawn from Figure~\ref{fig:apo_snr}.} 
    \label{fig:snr_noise}
\end{figure*}

In order to make mock 2MASS magnitudes and APOGEE ASPCAP outputs, we add noise to the corresponding simulated star properties and report a accompanying uncertainty. This procedure is described in Section \ref{sec:mock}, but here we provide further details on the derived empirical noise models.

In Figure \ref{fig:2mass_nosie}, we show the mean and standard deviation (shaded region) of the 2MASS catalog magnitude uncertainties as a function of magnitude for the J (blue), H (green), and $\rm{K_s}$ (pink) bands. These are calculated using 0.1 mag bins. In the bottom panel, we show the number of sources per bin as a function of magnitude. In general, we can see a jump in the average uncertainty at magnitudes $<$5 due to the 2MASS saturation limit. Between magntiudes of 5 to 10 we see the average uncertainty stays roughly constant at values between 0.02-0.06 mag. At magnitudes >10, we see the average uncertainty increase with increasing magnitude as expected.

Figure \ref{fig:apo_snr} we show the average and standard deviations of the APOGEE SNR for the DR19 GG data as a function H magnitude. Similar to Figure \ref{fig:2mass_nosie}, we use 0.1 mag bins. At H<7, the average SNR stays roughly constant at 250-300. These stars are brighter than the nominal APOGEE bright limit and are therefore targeted with a different strategy than the rest of the survey. At H>7, we see the average SNR decreases sharply with H magnitude. This is because all targets receive the same 15-minute exposure, and the observing strategy calls for a single epoch for the majority of them. Only a subset of stars have repeat observations.

The average and standard deviations of the ASPCAP stellar parameters and abundances as a function of APOGEE SNR for the DR19 GG data are shown in Figure \ref{fig:snr_noise}. For all the parameters and abundances we see the uncertainty decreases as the SNR increases. The $\rm{T_{eff}}$, log $g$, [M/H], [C/H] and [N/H] are calculated simultaneously through $\chi^2$ spectral template fitting. This is why [M/H], [C/H], and [N/H] have weaker relationships with SNR and larger standard deviations than the other elemental abundances. The average $\rm{T_{eff}}$ uncertainty ranges from 20 K at the low SNR end to 1 K at high SNR. The average log $g$ uncertainty ranges from 0.07 dex to 0.004 dex. [M/H] average uncertainty ranges from 0.001-0.02 dex, while the [N/H] average uncertainty is much larger ranging from 0.003-0.4 dex. The [C/H] average uncertainty is in the middle with values between 0.0025-0.07 dex. Of the individually measured elemental abundances, Fe is the most precise, with uncertainties of 0.001 to 0.01 dex. This is expected, as the Fe has the most lines and the fit uses almost an order of magnitude more spectral windows than any other atomic species. The ASPCAP abundance uncertainty floor is 0.001 dex, which causes the flattening in the [Fe/H] uncertainty at SNR>150.  For all the other elements (Mg, Si, S, O and Ca) the average uncertainty ranges from 0.025-0.002 dex. We note that these values encompass the noise driven statistical uncertainties but underestimate the total uncertainty because the systematic uncertainties are not taken into account \citep{Meszaros2025}.

\section{Chemical Abundances}
\begin{figure*}
    \centering
    \includegraphics[width=\linewidth]{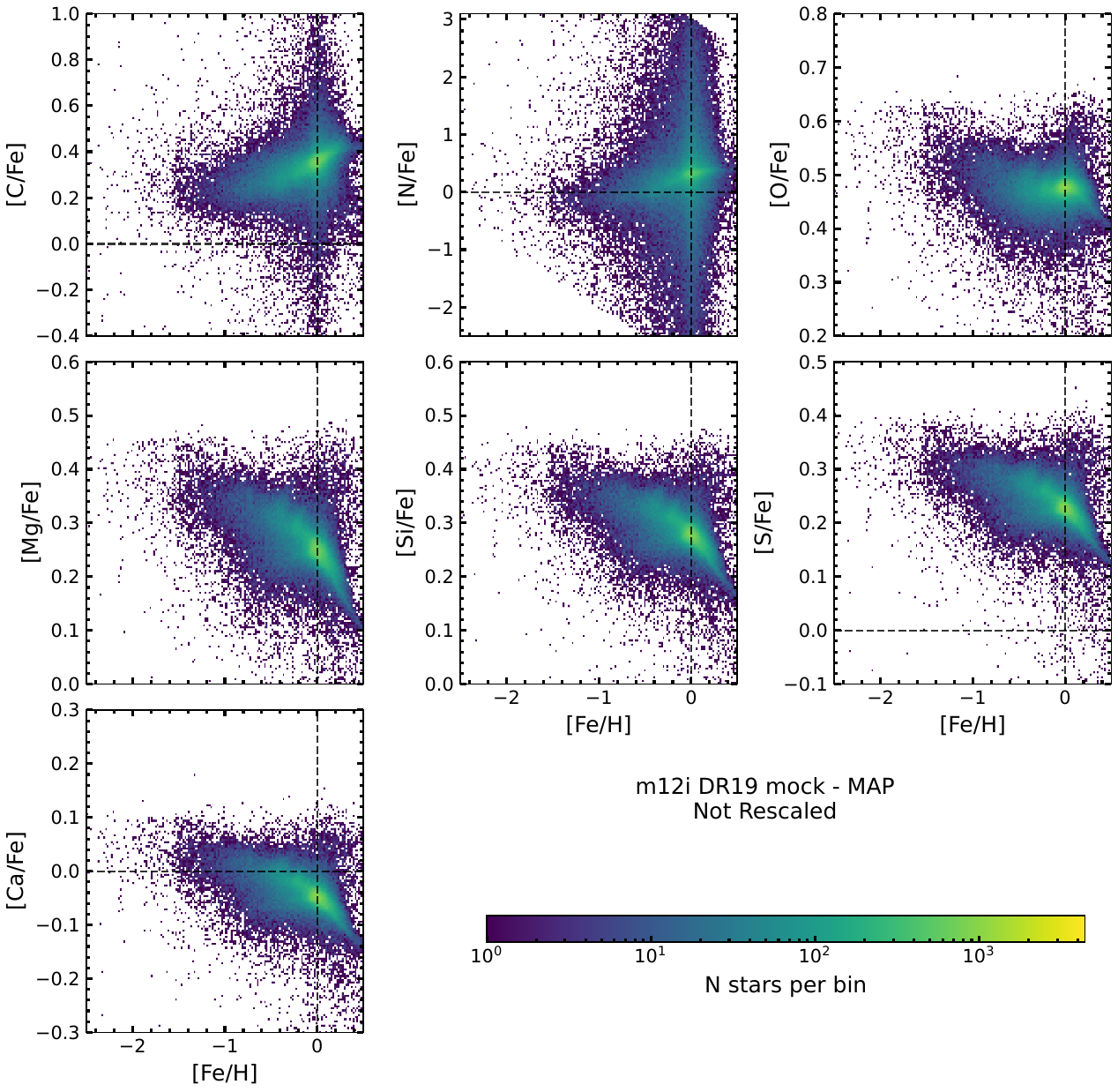}
    \caption{[X/Fe] versus [Fe/H] for the m12i  mock catalog after applying the DR19 selection function (MAP estimate). Each panel shows one element (C, N, O, Mg, Si, S, and Ca), computed from the mock's noisy, ASPCAP-like abundances as $[\mathrm{X/Fe}] = [\mathrm{X/H}] - [\mathrm{Fe/H}]$, with no rescaling. Colors show the number of stars per bin on a logarithmic scale. Dashed lines mark $[\mathrm{Fe/H}] = 0$ and $[\mathrm{X/Fe}] = 0$. For most elements the $[\mathrm{X/Fe}] = 0$ line sits near the bottom of the panel, or below  it. Near solar metallicity the mock [X/Fe] values lie about
    0.23--0.48~dex above solar for C, N, O, Mg, Si, and S, and about
    0.05~dex below solar for Ca. }
    \label{fig:abund}
\end{figure*}

\begin{figure*}
    \centering
    \includegraphics[width=\linewidth]{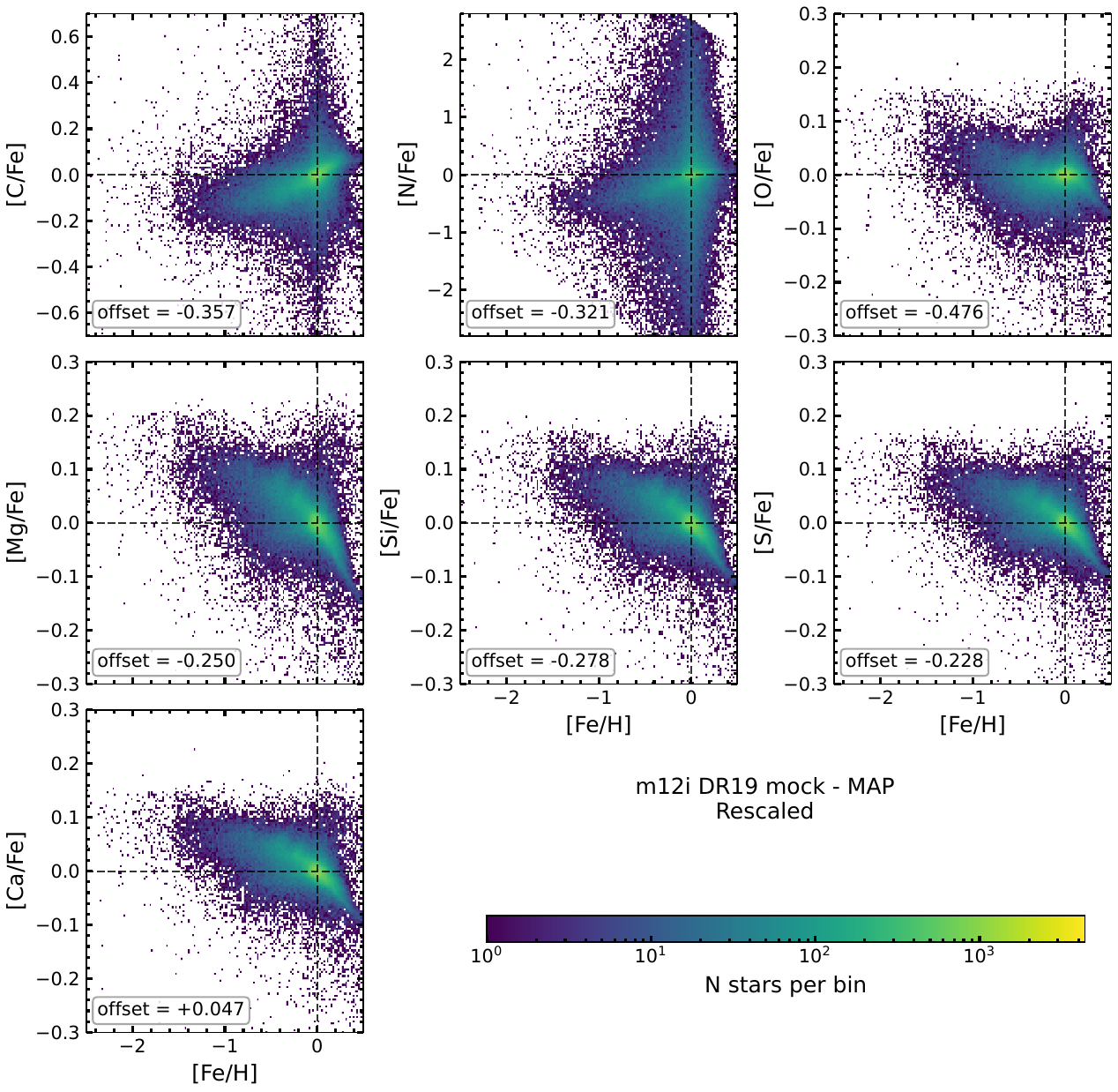}
    \caption{Same as Figure~\ref{fig:abund}, but with each
    element's [X/Fe] shifted so that the median [X/Fe] of stars with
    $|[\mathrm{Fe/H}]| < 0.05$ is zero. The shift added to each element is
    printed in the lower-left corner of its panel. }
    \label{fig:abund_scale}
\end{figure*}

The abundances included in the mock catalog are the seven elements (C, N, O, Mg, Si, S, and Ca)  that the FIRE-2 simulations track that are also included in the APOGEE ASPCAP catalogs. The [X/Fe] ratios for these elements are shown as a function of metallicity in Figure \ref{fig:abund} for the nominal mock catalog that uses the MAP selection function. The larger uncertainties for [C/Fe] and [N/Fe] (see Figure \ref{fig:snr_noise}) cause significant spread in the abundances for these elements. The reasons that the FIRE-2 abundances do not match the MW are complex, from limitations to the input yields and delay-time distributions for different channels of enrichment \citep[][Gandhi et al. in prep]{Hopkins2018b, 2020MNRAS.492.3465H, 2021MNRAS.508..508M,Gandhi2022, 2023MNRAS.519.3154H} to the method of implementation of metal transport through diffusion \citep{2018MNRAS.474.2194E} to the choice to initialize gas metallicity at a set value rather than model Pop III stars directly \citep[][Meizani et al. in prep]{Hopkins2018b,2023MNRAS.519.3154H}. To bring the mock abundances closer to the MW, we rescale the [X/H] values so that the median [X/Fe]=0 between -0.5 < [Fe/H] <+0.5. This is not necessarily true in the MW, but it is true for the solar neighborhood, and this change brings the abundances closer to MW trends in general. The rescaling is unlikely to be appropriate for stars with abundances very far from solar given the complexity of the modeled enrichment processes, but at least allows for a more informative comparison near the Solar circle. 

Figure \ref{fig:abund_scale} is the same as Figure \ref{fig:abund}, but with the rescaled abundances. The value that each are rescaled by is shown in the lower left corner. Ca is the only element that gets scaled with a positive value in that the abundances needed to increase. This is also the smallest shift with a value of 0.047 dex. The largest shift is -0.476 dex applied to the O abundances. The average shift is -0.27 dex.

\section{Milky Way Dust Comparison}

\begin{figure*}
    \centering
    \includegraphics[width=\linewidth]{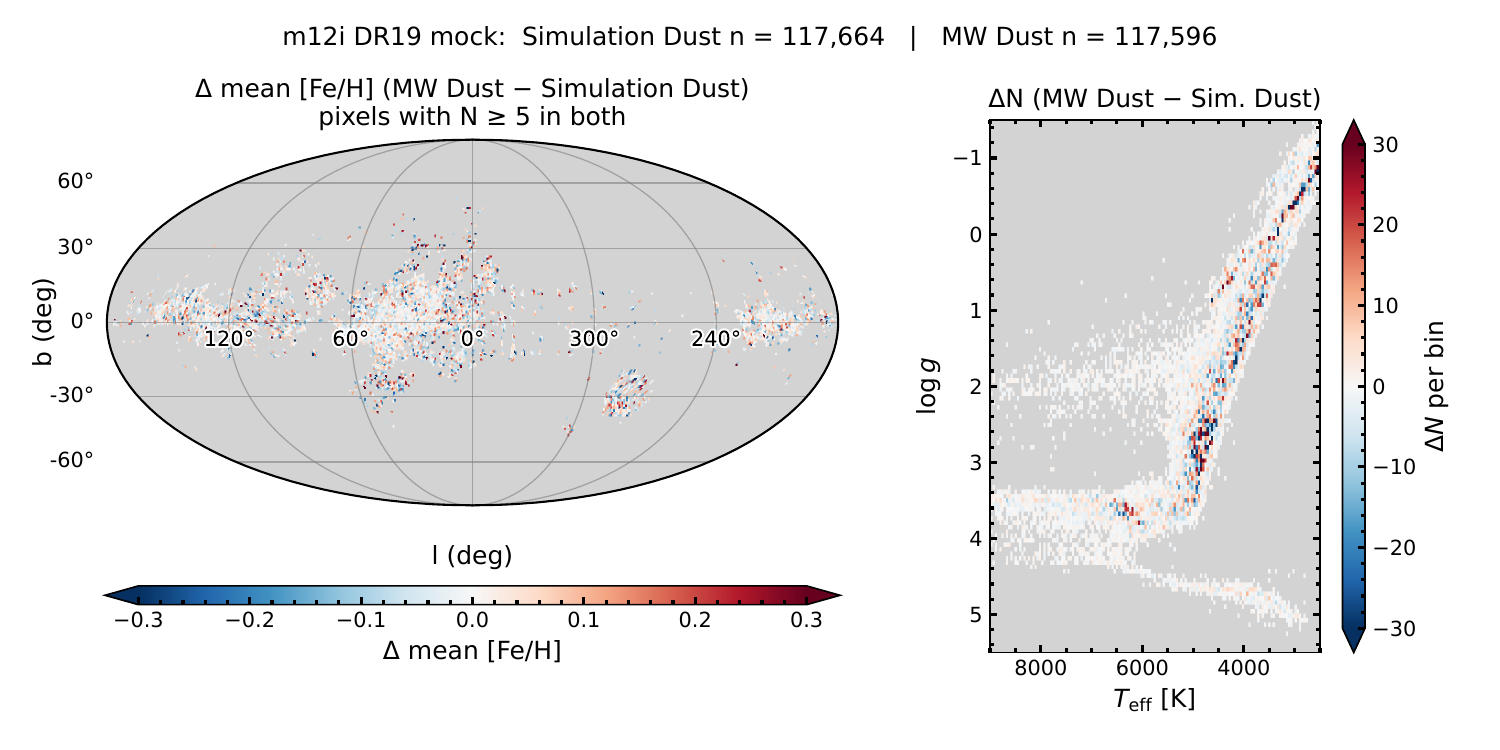}
    \caption{Effect of the dust prescription on the m12i mock catalog after applying the SDSS-V DR19 GG selection function (MAP estimate). We compare the mock built with the simulation's dust model (``Simulation Dust''; $N = 118{,}088$ stars) to the same mock built with extinction from the three-dimensional MW dust map (``MW Dust''; $N = 118{,}256$ stars). In both panels, differences are MW Dust minus Simulation Dust, and all quantities are the mocks' true (noise-free) stellar parameters. The left panel shows the difference in mean [Fe/H] per pixel, shown in Galactic coordinates. Only pixels containing at least five stars in both catalogs are shown. The median absolute difference is 0.05\,dex, and no coherent large-scale pattern is visible. The right panels shows the difference in star counts per bin in the Kiel diagram. Red (blue) bins contain more (fewer) stars in the MW Dust catalog. The largest differences ($|\Delta N| \sim 20$--$30$) occur in the densest regions, such as the red clump, where they are comparable to the Poisson scatter ($\sqrt{N} \approx 30$ for $N \sim 10^3$ stars per bin). }
    \label{fig:dust_diff}
\end{figure*}

For the nominal version of the catalog we prioritize self-consistency and calculate the dust extinction based on the metal-weighted hydrogen column density along the line of sight to each mock star (see Section \ref{sec:mock}). However, we also provide a version where we utilize the MW 3D dust map \citep{Green2019,Zucker2025}, as described in Section \ref{sec:mock}. Since the H and G-H magnitudes change with the dust treatment, applying the selection function yields slightly different stellar samples in each mock. We see the MW dust extinction keeps more stars within the survey observing strategy with a total of 118,256 stars in the final catalog, while the nominal catalog using the simulation-based dust estimate has 118,088 stars.

We compare the mocks with different dust treatment in Figure \ref{fig:dust_diff}. In the left panel, we show the mean metallicity difference as a function of sky position between the MW dust and simulation dust catalogs. While some pixels change as much as 0.3 dex, there are generally no significant trends. The right panel shows the difference in number of stars between the catalog as a function of $\rm{T_{eff}}$ and log $g$. We can see the largest differences are mostly for cool giants. Because of the APOGEE magnitude  limits, these stars are generally more distant than the dwarfs, including in the inner Galaxy where dust extinction is highest. Therefore, they are the most likely to be effected by changes in the dust extinction treatment. We also confirm there are no significant differences in the azimuthal structure between the mocks with different dust treatment.

\section{FIRE-2 Mock Variations}

\begin{figure*}
    \centering
    \includegraphics[width=.8\linewidth]{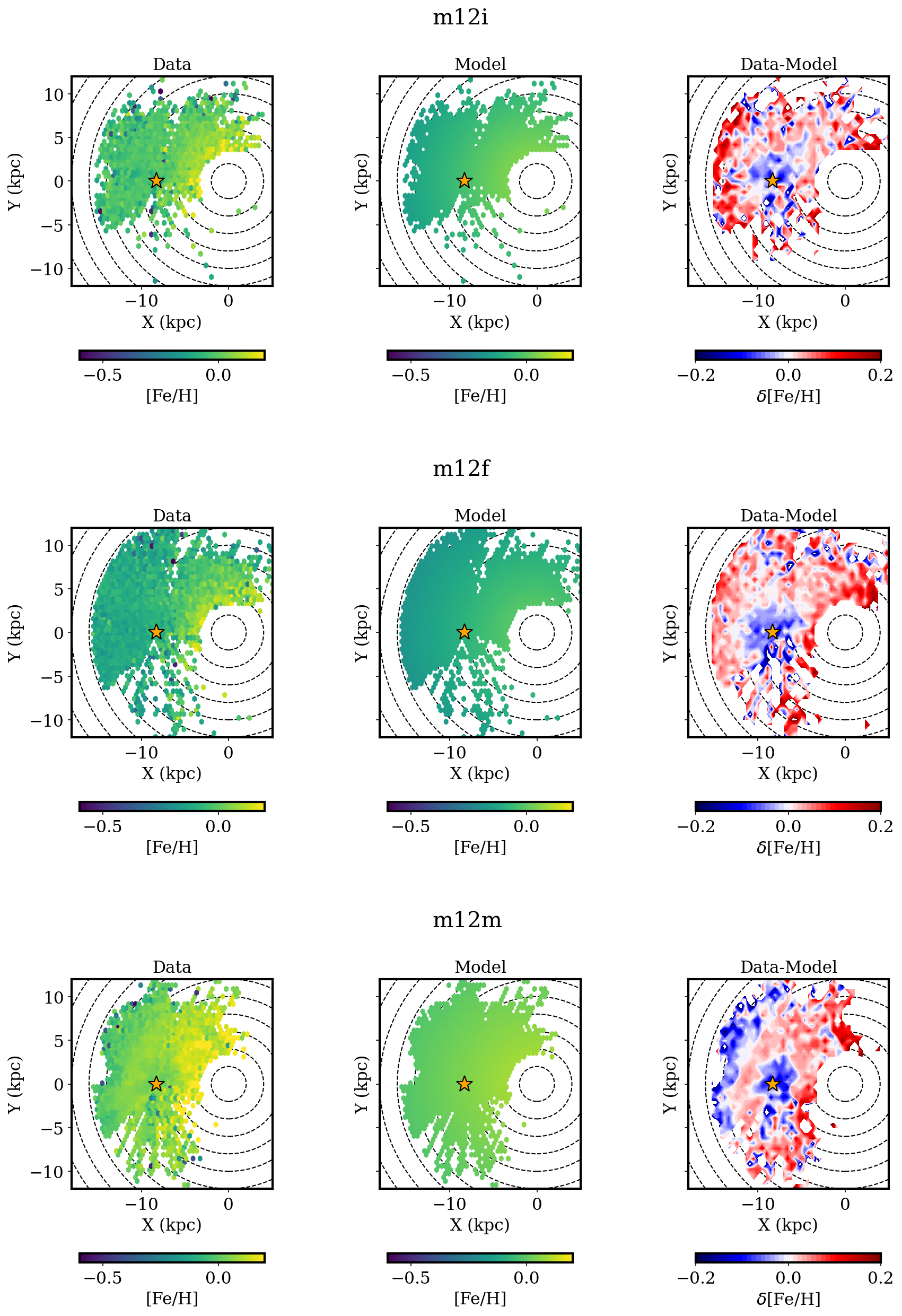}
    \caption{Same as Figure \ref{fig:azvars}, but with the addition of the 2 other FIRE-2 galaxies (m12f and m12m) for which we computed mock catalogs. Each galaxy shows distinct azimuthal variations in average metallicity, but they all have the solar neighborhood as below average metallicity. M12m has the most stellar mass, is the most metal-rich and shows the most azimuthal structure in average metallicity.   }
    \label{fig:az_var_varis}
\end{figure*}

In addition to the nominal m12i  mock catalog, we calculate mock catalogs for two other FIRE-2 galaxies which have $\textit{Gaia}$ DR3 mocks presented in \citet{Nguyen2024}. These are all calculated using the MAP estimate of the selection function and using the dust model from the simulations, rather than using a MW dust map. 

One of the largest differences between the catalogs are the total number of stars. M12m has the highest stellar mass ($M_{*}=10\times10^{10} M_{\odot}$), then m12f ($M_{*}=6.9\times10^{10} M_{\odot}$) and m12i has the lowest ($M_{*}=5.3\times10^{10} M_{\odot}$) \citep{McCluskey2024}. As the selection function draws a fraction of stars from the simulated galaxy, the mock catalog size scales with the stellar mass of the simulated galaxy. The m12m mock has a total of 337,763 stars, while m12f has 199,284 stars and m12i has 118,088 stars. 

In order to briefly explore the differences between these catalogs, we remake Figure \ref{fig:azvars}, but with the other mock galaxies. The top row of Figure \ref{fig:az_var_varis} is identical to the bottom row of Figure \ref{fig:azvars}, with the middle and bottom rows showing the equivalent for m12f and m12m, respectively. Consistent with expectations for the highest stellar mass, m12m is also the most metal-rich and the average metallicity of stars in the mock catalog is -0.03 dex. Although, m12f has slightly more stellar mass, it is a little bit more metal-poor than m12i where the average metallicity of stars in the mock for m12f is -0.14 dex while the same for m12i is -0.13 dex. There are also difference in the radial metallicity gradient as measured from the mocks. M12i has the strongest radial metallicity gradient (-0.0156 dex/kpc) which is most similar to the estimate from the SDSS-V DR19 GG observed sample (-0.0572 dex/kpc). Therefore, we adopt it as our nominal mock catalog in this work. M12m has the weakest radial metallicity gradient at -0.0098 dex/kpc. 

In the right panels of Figure \ref{fig:az_var_varis}, we can see that the azimuthal variations in the mean metallicities varies between the FIRE-2 galaxies. M12m especially appears to show spiral structure-like variations, but further investigation that is beyond the scope of this work is needed to confirm this. Although each galaxy shows unique variations, in all cases the solar neighborhood is more metal-poor than expected from the radial metallicity gradient, while other azimuths at the same radius are more metal-rich. This is likely induced by the selection function, as supported by the absence of this feature in the uniform mock (Figure \ref{fig:azvars}). In contrast, the MW data show a relatively metal-rich solar neighborhood. However, we cannot assume the selection function has the same effect on the MW data, since it depends on the underlying stellar populations. Further work modeling the MW's azimuthal substructure while accounting for the selection function is required to fully understand this signal and its effect on MW measurements.

\section{The \pkgname\ Package}\label{app:pack}

The method of Section~\ref{sec:sf} is implemented in the Python package \pkgname\footnote{\url{\pkgurl}; documentation at \url{\pkgdocs}.}\citep{sfpackage}. In this paper we use it in a single configuration: the selection function of the DR19 GG sample as a function of sky position, $H$, and $G-H$, reported as the map and mean, and with a minimum of 20 2MASS sources per cell, which we then apply to the FIRE mock catalogs in Section~\ref{sec:mock}. The package itself is more general. In this appendix we summarize its capabilities (Appendix~\ref{app:capabilities}) and give a short example of its use (Appendix~\ref{app:code}).

\subsection{Capabilities}\label{app:capabilities}

The only input the user provides is a table of observed stars with sky positions, $H$ magnitudes, and, optionally, Gaia $G$ magnitudes. The 2MASS denominator, which is the expensive part of the calculation, ships with the package and contains $5.46\times10^{8}$ 2MASS sources counted on a HEALPix $N_{\rm side}=64$ grid in $0.5$,mag bins over $-3 \le H < 18$ and $-2.5 \le G-H < 14.5$. Because the observed sample is supplied by the user, the selection function can be computed for any SDSS-V APOGEE program, data release, or subset defined by quality or stellar-parameter cuts, not only for the sample considered here. However, it is important to note that the underlying assumption that 2MASS is complete begins to break down at H>12 towards the inner Galaxy so selection functions calculated with data beyond that cut are likely overestimating the completeness. The user can also choose (i) a selection function in $H$ alone or in both $H$ and $G-H$; (ii) the magnitude range and bin widths, in multiples of $0.5$\,mag;(iii) the minimum number of 2MASS sources per cell before pixels are merged, and the coarsest HEALPix resolution allowed; and (iv) whether each cell reports the posterior mean,
$(N_{\rm obs}+1)/(N_{\rm 2MASS}+2)$, or the posterior mode,
$N_{\rm obs}/N_{\rm 2MASS}$. Building a selection function for the full DR19 GG sample ($\approx2.4\times10^{5}$ stars) takes seconds on a laptop, so it is practical to rebuild it for each subsample in an analysis. The result can be saved to and reloaded from disk, and its \texttt{query} method returns the selection probability for any set of sky positions and magnitudes. This is what is needed to forward model a survey, whether to draw an observed sample from a model or simulation, as we do for the mock catalogs, or to include the selection in the likelihood when fitting a model to the observed stars.

\subsection{Example Usage}\label{app:code}

Below we compute the selection function for the DR19 GG sample from public data. A longer version of this example, including maps of the selection function, is provided as a tutorial notebook in the repository. The package is installed with
\begin{lstlisting}[language=bash]
pip install git+https://github.com/mlucey/sdssv_apogee_sf
\end{lstlisting}
and the 2MASS denominator is included in the installation. The observed sample is taken from the DR19 ASPCAP summary file\footnote{\url{https://data.sdss.org/sas/dr19/spectro/astra/0.6.0/summary/astraAllStarASPCAP-0.6.0.fits.gz}}. We select GG targets with the SDSS-V targeting flags and keep one entry per 2MASS source, matching how the denominator is counted.

\begin{lstlisting}
import numpy as np
from astropy.table import Table, unique
from sdss_semaphore.targeting import TargetingFlags

dr19 = Table.read('astraAllStarASPCAP-0.6.0.fits.gz', hdu=2)
flags = TargetingFlags(dr19['sdss5_target_flags'])
gg = flags.in_program('mwm_galactic') | flags.in_program('mwm_gg')
obs = unique(dr19[gg], keys='obj', keep='first')   # one row per 2MASS star

h = np.asarray(obs['h_mag'], dtype=float)
g = np.asarray(obs['g_mag'], dtype=float)
h[(h < -100) | (h > 20)] = np.nan                  # remove sentinel values
g[(g <= 0) | (g > 22)] = np.nan
observed = Table({'ra': obs['ra'], 'dec': obs['dec'],
                  'h_mag': h, 'g_mag': g})
\end{lstlisting}

The selection function is then built from this table. The settings below match those used for the mock catalogs in Section~\ref{sec:mock}; omitting \texttt{g\_col} would instead give a selection function in $H$ only.

\begin{lstlisting}
from sdssv_apogee_sf import APOGEESelectionFunction

sf = APOGEESelectionFunction.from_observed(
    observed, ra_col='ra', dec_col='dec',
    h_col='h_mag', g_col='g_mag',   # selection in H and G-H
    estimator='map',                # posterior mode, N_obs / N_2MASS
    min_count=20, nside_min=8)      # pixel-merging settings
sf.write('gg_sf_dr19.npz')          # reload with APOGEESelectionFunction.read
\end{lstlisting}

Finally, \texttt{query} evaluates the selection function at any sky positions and magnitudes (scalars or arrays), returning \texttt{NaN} outside its range:

\begin{lstlisting}
from astropy.coordinates import SkyCoord

c = SkyCoord(l=30.0, b=5.0, unit='deg', frame='galactic')
S = sf.query(c, H=10.0, GH=3.0)   # probability that such a star was observed
\end{lstlisting}


\bsp	
\label{lastpage}
\end{document}